\documentclass[table]{ccjnl}
\usepackage{lipsum,amsmath}
\usepackage{cuted}
\usepackage{bm}
\usepackage{multirow}
\usepackage{soul, color, xcolor}
\soulregister{\cite}7 
\soulregister{\citep}7 
\soulregister{\citet}7 
\soulregister{\ref}7 
\soulregister{\pageref}7 
\graphicspath{{figures/}}

\title{3D Spectrum Mapping and Reconstruction under Multi-Radiation Source Scenarios}
\author{Jie Wang\inst{1,2}, Zhipeng Lin\inst{1,2}, Qiuming Zhu\inst{1,2,*}\corinfo{zhuqiuming@nuaa.edu.cn}, Qihui Wu\inst{1,2}, Tianxu Lan\inst{1,2}, Yi Zhao\inst{1,2}, Yunpeng Bai\inst{1,2}, Weizhi Zhong\inst{2,3}}
\receiveddate{***}
\reviseddate{***}
\Editor{***}

\address[1]{The Key Laboratory of Dynamic Cognitive System of Electromagnetic Spectrum Space, Nanjing University of Aeronautics and Astronautics, 211106, Nanjing, China}
\address[2]{College of Electronic and Information Engineering, Nanjing University of Aeronautics and Astronautics, 211106, Nanjing, China}
\address[3]{The Key Laboratory of Dynamic Cognitive System of Electromagnetic Spectrum Space, College of Astronautics, Nanjing University of Aeronautics and Astronautics, Nanjing 211106, China}

\begin{document}

\maketitle

\begin{abstract}
Spectrum map construction, which is crucial in cognitive radio (CR) system, visualizes the invisible space of the electromagnetic spectrum for spectrum-resource management and allocation. Traditional reconstruction methods are generally for two-dimensional (2D) spectrum map and driven by abundant sampling data. In this paper, we propose a data-model-knowledge-driven reconstruction scheme to construct the three-dimensional (3D) spectrum map under multi-radiation source scenarios. We firstly design a maximum and minimum path loss difference (MMPLD) clustering algorithm to detect the number of radiation sources in a 3D space. Then, we develop a joint location-power estimation method based on the heuristic population evolutionary optimization algorithm. Considering the variation of electromagnetic environment, we self-learn the path loss (PL) model based on the sampling data. Finally, the 3D spectrum is reconstructed according to the self-learned PL model and the extracted knowledge of radiation sources. Simulations show that the proposed 3D spectrum map reconstruction scheme not only has splendid adaptability to the environment, but also achieves high spectrum construction accuracy even when the sampling rate is very low.
\end{abstract}
\section{INTRODUCTION}
\label{INTRODUCTION}
\begin{figure*}[htbp]
	\centering
	\includegraphics[width=1\textwidth]{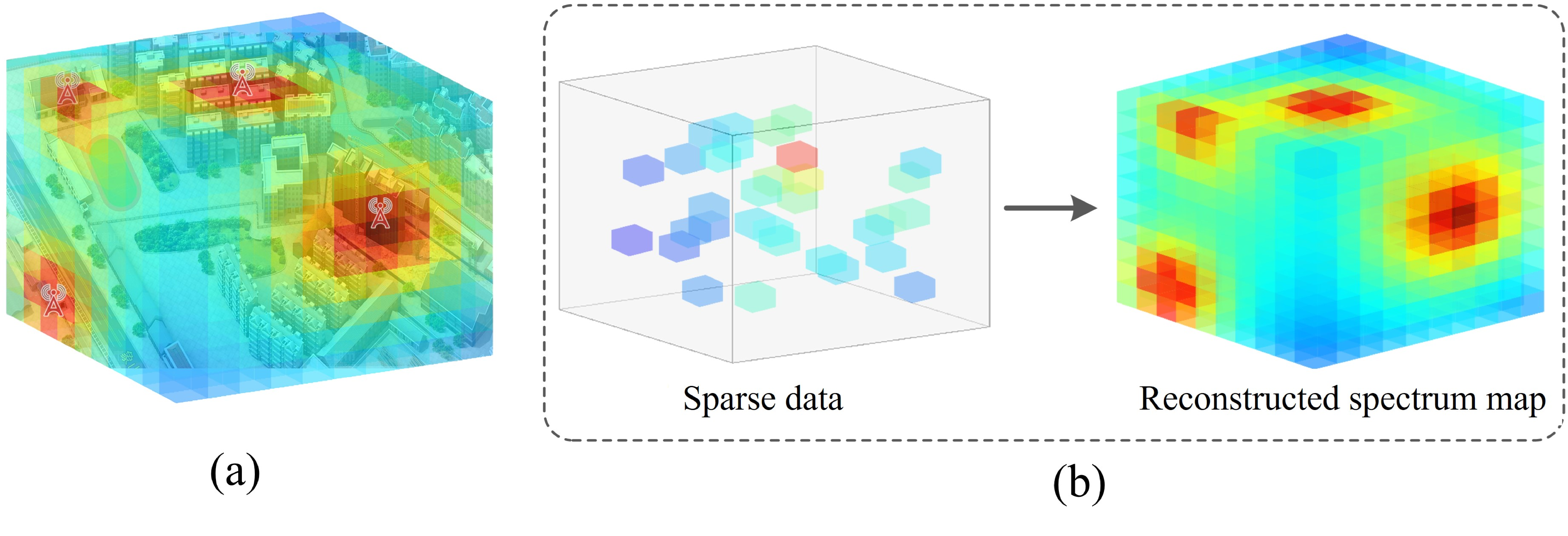}
	\caption{(a): Realistic 3D spectrum map under an urban scenario; (b): Reconstruction model of 3D spectrum map.}
	\label{fig1}
\end{figure*}
With the extensive growth of wireless frequency devices, the problem of electromagnetic spectrum resource deficit is increasing severe. In this case, the cognitive radio (CR) technique is proposed, which is beneficial for efficient management and allocation for rational spectrum resources \cite{1,9,15,17,26,34,39,40}. A key task in CR is using spectrum awareness to accurately determine the spectrum situation, including the spectrum environments, the received signal strength, the spectrum busy or the idle states and the spectrum access protocols, etc \cite{4,27,28,31,32}. Radio environment map (REM), also known as spectrum map, comprising time, frequency, signal strength, location and other multi-dimensional spectrum information can visualize spectrum state awareness results on a geographical map \cite{5,35}. REM is useful for numerous applications in wireless communication, e.g., opportunistic spectrum access (OSA), spectrum management, and interference coordination \cite{4,2,25,30,36}. However, considering the limited capability of sensing devices and the costs of measurement, it is necessary to construct an accurate and robust REM with limited sampling data.

Many spectrum map reconstruction methods have emerged in recent years. In \cite{18}, the authors constructed a REM to process the data collected by heterogeneous spectrum sensors. A novel machine learning approach based on generative adversarial networks (GANs) was presented in \cite{19}, which considers the completion of spectrum map from the perspective of image processing. The authors in \cite{20} converted the estimation task into an image reconstruction task by image color mapping. Nevertheless, these methods only focus on two-dimensional (2D) ground spectrum mapping, which cannot meet the needs of three-dimensional (3D) spectrum monitoring. Recently, electromagnetic spectrum monitoring is facing a huge challenge that the monitoring space extends from the land to the air. There have been many high-flexibility air spectrum sensing devices, such as unmanned aerial vehicle (UAV), which have excellent spatial freedom and abundant spectrum access opportunities. Compared to fixed ground sensing devices, they can more easily capture the dynamic changes of spectrum \cite{22}, and can be used to construct the spectrum map of 3D spaces.

Typically, spectrum map reconstruction can be divided into data-driven methods and model-driven methods. The data-driven methods are usually regarded as a matrix completion problem by interpolation methods. In \cite{6}, the authors proposed an inverse distance weighted (IDW) interpolation construction method. It does not take any physical fact into account, degrading the performance in terms of REM construction of urban environments with severe shadow fading. Kriging interpolation is another spatial interpolation method \cite{33}. In \cite{7}, the authors proposed an optimal data-driven spectrum mapping method based on distributed Kriging. Moreover, by extending the low-rank property from matrices to tensors, a tensor completion method combining the prediction model was proposed in \cite{8}. However, these methods can only achieve satisfactory performance when a large amount of data is collected, and they cannot well reflect the essence of electromagnetic propagations. In addition, sampled locations have a great influence on the completion performance. If the sampled location is far away from the radiation source, the completion performance will be poor.

Model-driven methods based on radiation source information and electromagnetic propagation model can be addressed in \cite{10,11}. They can achieve high mapping accuracy and are applicable to sparse sampling data. The authors in \cite{10} proposed a LocatIon Estimation based (LIvE) algorithm. The spectrum map is reconstructed by estimating the location of primary user (PU) based on the RSS value. However, the power of PU and the PL model, are not easy to obtain. In \cite{11}, authors proposed an indirect model-driven reconstruction method by taking the location, power and antenna pattern. In \cite{24}, a least absolute shrinkage and selection operator (LASSO) method was proposed by using the PL model for each radiation source. Nevertheless, these model-driven methods are not applicable to the sce- narios with multi-radiation sources.

In this paper, we propose a data-model-knowledge-driven 3D spectrum construction method under multi-radiation source scenarios. We first extract the knowledge of radiation sources in the region of interest (ROI) based on the sampling data, and then combine it with the propagation model to build a spectrum map that truly presents the real-time spectrum situation, which combines sampling data, propagation model and radiation source knowledge to achieve spectrum mapping.

The main contributions of this paper are summarized as follows: 
\begin{itemize}
	
	\item A maximum and minimum path loss difference (MMPLD) clustering algorithm is developed to detect the number of multi-radiation sources in the 3D space. Based on the propagation model, the clustering algorithm identifies the number of radiation sources in the measurement area, and thus, can achieve the spectrum map construction of 3D multi-radiation source scenarios.
	
	\item The shuffled frog leaping algorithm (SFLA) is proposed to estimate the high-dimensional parameters of the locations and the powers of radiation sources. SFLA can efficiently execute the process of multi-source fusion RSS data and achieve the parameters estimation of radiation sources.
	
	\item  Considering the dynamic nature of the urban environment, we self-learn the propagation model by modifying the PL model with sparse sampled data. The model self-learning algorithm can precisely capture the characteristics of electromagnetic propagation in real time.
	
\end{itemize}

The remainder of this paper is organized as follows. Section II introduces and elaborates the 3D spectrum reconstruction model. In Section III, a MMPLD-SFLA-based self-learning reconstruction mechanism of 3D spectrum map under multi-radiation source scenarios is proposed. Then, Section IV presents the simulation results and analyses. Finally, Section V draws conclusions. 
\section{SYSTEM MODEL}
\label{SYSTEM}
In this paper, we consider to construct the 3D spectrum map of multi-radiation source urban scenario as shown in Figure~\ref{fig1}. In the considered urban scenario, there are multi-radiation sources that have high priorities or legally licenses to utilize the frequency bands. Users opportunistically access the available spectrum imperceptibly. We construct a complete 3D spectrum map for rational use of spectrum resources, so that we can obtain the spectrum occupation status in the ROI and achieve efficient and reliable dynamic spectrum access in the future work. 

To construct the 3D spectrum map, we first discretize the whole ROI into small cubes. Each cube is colored according to the RSS of its central position, where red cubes represent high RSS values and blue cubes represent low RSS values. We establish a three-dimensional cartesian coordinate system according to the starting and ending points of the ROI and set up a spectrum tensor $\chi  \in {\Re ^{{N_1} \times {N_2} \times {N_3}}}$, in which ${N_1}$, ${N_2}$ and ${N_3}$ indicate the grid number in $x,y,z$ dimensions, respectively.

We proceed to mark the variables to be used. Note that it is impossible to deploy a large number of sensors or spectrum sensing devices for all cubes of the spectrum map due to high cost. Therefore, we complete the spectrum map by sampling data in the ROI as shown in Figure~\ref{fig1}. Suppose there are $N$ sampled points in the ROI, and each of them can provide its location and RSS values. $M$ radiation sources are located at ${\bf{t}}_j^t = \left[ {x_{_j}^t,y_{_j}^t,z_{_j}^t} \right],j = 1,2, \cdots M$ with the transmission power ${P_j}^t$. Let ${{\bf{T}}^r} \buildrel \Delta \over = \left[ {{\bf{t}}_{_1}^r,{\bf{t}}_{_2}^r, \cdots ,{\bf{t}}_{_N}^r} \right]$ be the collection of the sampling positions. ${\bf{t}}_{_i}^r = \left[ {x_{_i}^r,y_{_i}^r,z_{_i}^r} \right],i = 1,2, \ldots ,N$ are the coordinates of the cube in $\chi $. ${{\bf{P}}^r} \buildrel \Delta \over = \left[ {{P_1}^r,{P_2}^r, \cdots ,{P_N}^r} \right]$ is the collection of the sampled RSS with the value ${{\bf{T}}^r}$. We define $\mathbb{R} = \left\{ {{{\bf{R}}_1},{{\bf{R}}_2}, \ldots ,{{\bf{R}}_N}} \right\}$ as a set of samples, where ${{\bf{R}}_i} = ({\bf{t}}_i^r,P_i^r)$.
\section{3D SPECTRUM MAPPING AND RECONSTRUCTION}
\label{3D}
\subsection{The Procedure of 3D Spectrum Map }
\label{Overview}
\begin{figure*}[htbp]
	\centering
	\includegraphics[width=0.8\textwidth]{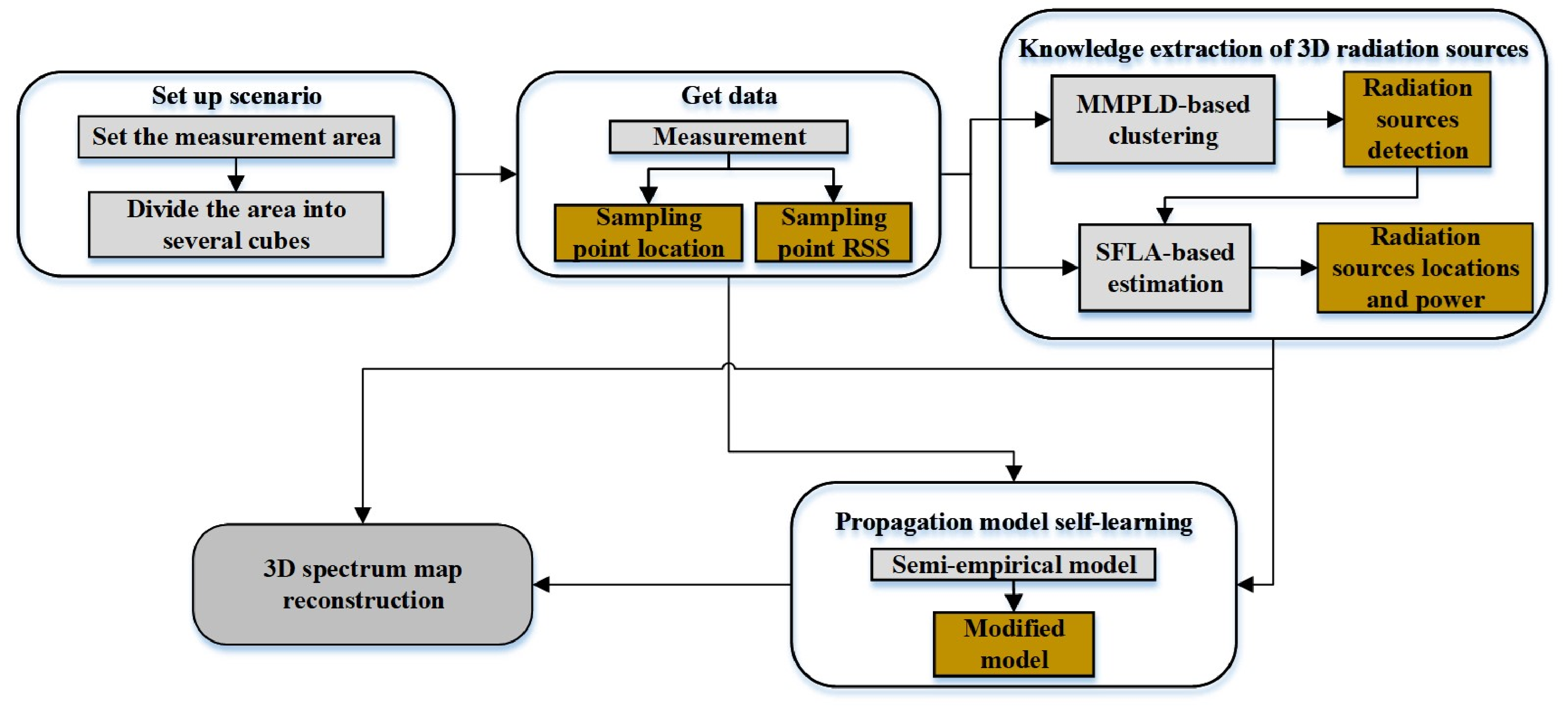}
	\caption{The flowchart of 3D spectrum map reconstruction.}
	\label{fig2}
\end{figure*}
Our 3D spectrum map reconstruction consists of four steps: $i)$ Scenario setting up and ROI modeling; $ii)$ Sampling data collection; $iii)$ Knowledge extraction of the radiation sources; $iv)$ Self-learning of the propagation model and reconstruction of the spectrum map. At the first step, we set up and model the ROI. We collect sampling data from the spectrum sensing devices within the ROI. We then design the MMPLD-clustering algorithm and SFLA-optimization algorithm to extract the knowledge from the ROI, including the total number of radiation sources, and the location and power of each radiation source. Finally, given the complexity of urban environment, we learn the PL model based on sampling data to make the algorithm have self-learning ability. The spectrum map is constructed based on the data, model and knowledge. The flow chart of the whole algorithm is shown in Figure~\ref{fig2}.
\subsection{3D Radiation Sources Number Detection}
\label{3D Radiant}
The challenge of constructing 3D multi-radiation source spectrum map is to mine the knowledge of the radiation sources. In this section, we first briefly review the $K$-means clustering algorithm, and then, design the MMPLD-based clustering algorithm to detect the number of radiation sources.
\subsubsection{Review of $K$-means Clustering Algorithm}
\label{Overview of $K$-means Clustering Algorithm}
Clustering data based on the measurements of similarity is critical in data process and pattern recognition. The basic idea of clustering data is to learn a set of clustering centers from the data to minimize the sum of squared errors between the data points and their nearest centers \cite{23}.

$K$-means is a classic clustering technique based on geometric centers and partitions \cite{37,38}. It is based on Euclidean distance and considers that the closer the distance between two objects is, the greater the similarity is. However, the initial clustering centers are difficult to be determined and the $K$ value needs to be known in advance. Therefore, $K$-means usually needs to be rerun multiple times with different initializations to find a good solution. The clustering algorithm based on the maximum and minimum distance (MMD) criterion, which is developed from $K$-means, selects the farthest point to be the clustering center. The MMD algorithm can dynamically determine clustering number and avoid the shortcomings of $K$-means \cite{12}. Note that a hyperparameter threhold can be set determine whether the clustering process needs to be continued. Nevertheless, the selection of the hyperparameter threhold directly affects the partition results. Traditional clustering algorithms takes the Euclidean distance as the basis, but it is not suitable for spectrum data processing. Moreover, they cannot accurately determine the number of clusters.
\subsubsection{MMPLD-based 3D Radiation Sources Number Detection}
\label{MMPLD-based 3D Radiant Number Detection}
Traditional clustering algorithms need to know the number of clusters in advance. However, in our paper, the number of clusters is the parameter that we need to estimate. The maximum and minimum distance (MMD) clustering algorithm, which is developed from K-means, can dynamically determine clustering number. Considering the traditional the Euclidean distance clustering criterion is not suitable for spectrum data, we propose a maximum and minimum path loss difference (MMPLD) criterion by exploiting the correlation of spectrum data. Besides, a criterion function is designed to determine the number of clusters at the end of the algorithm to balance the complexity of parameter settings. Given the electromagnetic propagation law, the RSS of a particular position is more consistent with the propagation law of the nearest radiation source than other radiation sources. The main procedure of the proposed MMPLD algorithm includes three steps: $i)$ The path loss difference (PLD) is calculated as the basic criterion; $ii)$ The selection of clustering centers and data classification are executed by maximum and minimum PLD criterion; $iii)$ The convergence condition is designed for the algorithm by calculating the criterion function of clustering number $K$ to determine the optimal clustering number $K$. 

The PL model of a free space can be expressed as
{\setlength\abovedisplayskip{0.2cm}
\setlength\belowdisplayskip{0.2cm}
\begin{equation}
{L_0} = 32.4 + 20{\log _{10}}\left( {{f_{\rm{c}}}} \right) + 20{\log _{10}}\left( d \right),
\label{eq1}
\end{equation}
where ${L_0}$ is the path loss in free space. $d$ and ${f_c}$ denote the distance between sampled position and radiation source and frequency, respectively. 

We define $\mathbb{C}$ as the set of the clustering centers. Firstly, we choose the ${{\bf{R}}_n} \in \mathbb{R}$, which has the highest $P_{^n}^r$ of ${{\bf{P}}^r}$ as the first clustering center, denoted by ${C_1}$, ${C_1} \in \mathbb{C}$. According to~\eqref{eq1}, we can obtain the theoretical PL value $L_i^{{C_1}},i = 1,2,3 \cdots ,N$ between ${C_1}$ and the $i - th$ sampling point. Collecting all the values, we have ${{\bf{L}}^{{C_1}}}$, as given by
\begin{equation}
{{\bf{L}}^{{C_1}}} = \left\{ {L_{^2}^{{C_1}},L_{^3}^{{C_1}}, \ldots ,L_{^N}^{{C_1}}} \right\}.
\label{eq2}
\end{equation}
${\bf{P}}_\Delta ^{{C_1}}$ collects all the difference RSS values between ${C_1}$ and all sampling points, which can be expressed as
\begin{equation}
\begin{array}{l}
{\bf{P}}_\Delta ^{{C_1}} = \left\{ {\Delta P_1^{_{{C_1}}},\Delta P_2^{_{{C_1}}}, \cdots ,\Delta P_N^{_{{C_1}}}} \right\}\\
= \left\{ \begin{array}{l}
\left| {{P_{{C_1}}}^r - {P_1}^r} \right|,\left| {{P_{{C_1}}}^r - {P_2}^r} \right|,\\
\ldots \left| {{P_{{C_1}}}^r - {P_N}^r} \right|
\end{array} \right\},
\end{array}
\label{eq3}
\end{equation}
where ${P_{{C_1}}}^r$ is the RSS of the first clustering center ${C_1}$. ${P_i}^r\begin{array}{*{20}{c}}
,&{i = 1,2, \ldots ,N}
\end{array}$ is the RSS value of the $i$-th sampling point, $\Delta P_i^{_{{C_1}}}$ is the absolute difference values of RSS between the first clustering center ${C_1}$ and the $i$-th sampling point, and $\Delta P_i^{_{{C_1}}}$ is the actual PL value. We define error set ${{\bf{\delta }}^{{C_1}}}$ as clustering criteria, according to the following formula:
\begin{equation}
\begin{array}{l}
{{\bf{\delta }}^{{C_1}}} = \left\{ {\delta _{_1}^{{C_1}},\delta _{_2}^{{C_1}}, \ldots ,\delta _{_N}^{{C_1}}} \right\}\\
= \left\{ \begin{array}{l}
\left| {L_1^{{C_1}} - \Delta P_1^{{C_1}}} \right|,\left| {L_2^{{C_1}} - \Delta P_2^{{C_1}}} \right|,\\
\cdots ,\left| {L_N^{{C_1}} - \Delta P_N^{{C_1}}} \right|
\end{array} \right\},
\end{array}
\label{eq4}
\end{equation}
where $\delta _{_I}^{{C_1}}$ is the PLD between the first clustering center ${C_1}$ and the $i $-th sampling point, and $\delta _{_I}^{{C_1}}$ is the absolute error between theoretical PL and actual PL. We choose the max value of ${{\bf{\delta }}^{{C_1}}}$ to be the second cluster center, denoted as ${C_2}$, as given by
\begin{equation}
{C_2} \leftarrow \max (d_1^{\min },d_2^{\min }, \ldots ,d_N^{\min }),
\label{eq5}
\end{equation}
where $d_i^{\min } = \delta _{_i}^{{C_1}},i = 1,2, \ldots ,N$ indicates the PLD between the point and the center of the class, to which the point belongs. We initiate $d_i^{\min }$ with $\delta _{_i}^{{C_1}}$ and its value is constantly updated as the clustering process progresses. To be specific, if $d_n^{\min } = \max (d_1^{\min },d_2^{\min }, \ldots ,d_N^{\min })$, the second cluster center ${C_2}$ is the $n$-th point.
Afterwards, we calculate $L_i^{{C_2}}$ according to~\eqref{eq1}. The ${\bf{P}}_\Delta ^{{C_2}}$ of the second cluster center ${C_2}$ are
\begin{equation}
\begin{array}{l}
{\bf{P}}_\Delta ^{{C_2}} = \left\{ {\Delta P_1^{_{{C_2}}},\Delta P_2^{_{{C_2}}}, \cdots ,\Delta P_N^{_{{C_2}}}} \right\}\\
= \left\{ \begin{array}{l}
\left| {{P_{{C_2}}}^r - {P_1}^r} \right|,\left| {{P_{{C_2}}}^r - {P_2}^r} \right|,\\
\ldots \left| {{P_{{C_2}}}^r - {P_N}^r} \right|
\end{array} \right\},
\end{array}
\label{eq6}
\end{equation}
where $\Delta P_i^{_{{C_2}}}$ is the actual PL between ${C_2}$ and the $i$-th sampling point. The PLD set ${{\bf{\delta }}^{{C_2}}}$ in the second class can be obtained by
\begin{equation}
\begin{array}{l}
{{\bf{\delta }}^{{C_2}}} = \left\{ {\delta _{_1}^{{C_2}},\delta _{_2}^{{C_2}}, \ldots ,\delta _{_N}^{{C_2}}} \right\}\\
= \left\{ \begin{array}{l}
\left| {L_1^{{C_2}} - \Delta P_1^{{C_2}}} \right|,\left| {L_2^{{C_2}} - \Delta P_2^{{C_2}}} \right|,\\
\cdots ,\left| {L_N^{{C_2}} - \Delta P_N^{{C_2}}} \right|
\end{array} \right\},
\end{array}
\label{eq7}
\end{equation}
where $\delta _{_I}^{{C_2}}$is the PLD between the second clustering center ${C_2}$ and the $i $-th sampling point. We update $d_i^{\min }$ by
\begin{equation}
d_i^{\min } = \min \left( {\delta _i^{{C_1}},\delta _i^{{C_2}}} \right),i = 1,2, \ldots ,N.
\label{eq8}
\end{equation}

According to $d_i^{\min }$, we categorize the $i$-th sampling point into the class with the least PLD. Specifically, if $d_i^{\min } = \delta _i^{{C_m}},m = 1,2$, then the $i$-th point is assigned to class $m$. Furthermore, the $q$-th clustering center ${C_q}$ is selected from the maximum value of $d_i^{\min }$, as given by
\begin{equation}
{C_q} \leftarrow \max (d_1^{\min },d_2^{\min }, \ldots ,d_N^{\min }).
\label{eq9}
\end{equation}
If $d_n^{\min } = \max (d_1^{\min },d_2^{\min }, \ldots ,d_N^{\min })$, ${C_q}$ is the $n - th$ point. We keep obtaining the PLD ${{\bf{\delta }}^{{C_q}}}$ of the sampling points in the $q - th$ class
\begin{equation}
\begin{array}{l}
{{\bf{\delta }}^{{C_q}}} = \left\{ {\delta _{_1}^{{C_q}},\delta _{_2}^{{C_q}}, \ldots ,\delta _{_N}^{{C_q}}} \right\}\\
= \left\{ \begin{array}{l}
\left| {L_1^{{C_q}} - \Delta P_1^{{C_q}}} \right|,\left| {L_2^{{C_q}} - \Delta P_2^{{C_q}}} \right|,\\
\cdots ,\left| {L_N^{{C_q}} - \Delta P_N^{{C_q}}} \right|
\end{array} \right\},
\end{array}
\label{eq10}
\end{equation}
where $\delta _{_I}^{{C_q}}$ is the PLD between the $q$-th clustering center ${C_q}$ and the $i$-th sampling point; $L_i^{{C_q}}$ is the theoretical path loss value between ${C_q}$ and the $i$-th sampling point in the $q$-th class, and $\Delta P_i^{_{{C_q}}}$ is the actual path loss between ${C_q}$ and the $i$-th sampling point.

Meanwhile, we update the $d_i^{\min }$ of each point:
\begin{equation}
d_i^{\min } = \min \left( {\delta _i^{{C_1}},\delta _i^{{C_2}}, \ldots ,\delta _i^{{C_q}}} \right),i = 1,2, \ldots, N.
\label{eq11}
\end{equation}
We reclassify the sampling points according to the $d_i^{\min }$ until the end condition of the algorithm is satisfied. If $d_i^{\min } = \delta _i^{{C_m}},m = 1,2, \ldots ,q$, then the $i$-th point is assigned to the class $m$. There is a criterion that we select the clustering centers according to the maximum PLD, and classify points according to the minimum PLD. The final clustering number $K$ is the number of radiation sources need to be detected.
The selection of $K$ depends on whether the current clustering status conforms to the maximum similarity within the cluster and the minimum similarity between the clusters. In order to determine the optimal value $K$, we define the criterion function $\varpi ( * )$ as the convergence condition of the algorithm.

For the $i$-th point, which is in class $\gamma $, we define the function $\varsigma (i)$ of the point with in other classes, given by
\begin{equation}
\varsigma (i) = \frac{1}{{\left( {K - 1} \right)}}\sum\limits_{{C_{\rm{q}}},{C_{\rm{q}}} \ne \gamma }^{K - 1} {\delta _i^{{C_q}}},
\label{eq12}
\end{equation}
where $\delta _i^{{C_q}}$ is the PLD calculated by~\eqref{eq15}, and it does not belong to the current class $\gamma $. Then for the number $K$ of clusters we can define the criterion function $\varpi (K)$, given by
\begin{equation}
\varpi (K) = \frac{1}{N}\sum\limits_{i = 1}^N {\frac{{d_i^{\min }}}{{\varsigma (i)}}},
\label{eq13}
\end{equation}
where $d_i^{\min }$ is the PLD between the $i$-th sampling point and the clustering center ${C_\gamma }$ of the class $\gamma $.

We set two hyperparameters as thresholds as the convergence conditions of the algorithm, given by
\begin{equation}
\left\{ \begin{array}{l}
\left| {(\varpi (K) - \varpi (K - 1))} \right| \le {\sigma _1}\begin{array}{*{20}{c}}
,&{\begin{array}{*{20}{c}}
	{{\rm{if}}}&{1 \le K < 3}
	\end{array},}
\end{array}\\
\left| {\frac{{(\varpi (K) - \varpi (K - 1))}}{{(\varpi (K - 1) - \varpi (K - 2))}}} \right| \le {\sigma _2}\begin{array}{*{20}{c}}
,&{\begin{array}{*{20}{c}}
	{{\rm{if}}}&{K \ge 3}
	\end{array},}
\end{array}
\end{array} \right.
\label{eq14}
\end{equation}
where ${\sigma _1}$ and ${\sigma _2}$ are the thresholds of the clustering algorithm. When $K$ satisfies the above inequality, the clustering algorithm ends and the $K$ value is the number of radiation sources.
\subsection{SFLA-based 3D Radiant Location}
\label{SFLA-based 3D Radiant Location}
In this section, we estimate the parameters of the radiation sources including locations and powers. Suppose the transmission power of the $j$-th radiation source is ${P_j}^t,j = 1,2, \cdots K$ and the location of the $j$-th radiation source is ${\bf{t}}_j^t = \left[ {x_{_j}^t,y_{_j}^t,z_j^t} \right],j = 1,2, \cdots K$. For each radiation source, we define a coefficient ${\eta _j}$ related to its propagation loss. On the basis of the electromagnetic propagation law, the RSS value ${\hat P_{ij}}^r$ of the $i$-th of $N$ sampling points from the $j$-th radiation source is
\begin{equation}
\begin{array}{l}
{{\hat P}_{ij}}^r = {\eta _j}{P_{\rm{j}}}^t{d_{ij}}^{ - \alpha },\\
i = 1,2, \ldots N\begin{array}{*{20}{c}}
,&{j = 1,2, \ldots K}
\end{array}
\end{array}
\label{eq15}
\end{equation}
where ${d_{ij}}$ is the Euclidean distance between the location ${\bf{t}}_j^r$ of the $i$-th sampling point and the location ${\bf{t}}_j^t$ of the $j$-th radiation source, and $\alpha $ is the path loss index.

For each sampled point, we can only measure the combined RSS of all radiation sources. Then its RSS value from all radiation sources are added up to generate the measured data ${\hat P^r}_i$, as given by
\begin{equation}
{\hat P_i}^r = \sum\limits_{j = 1}^K {{{\hat P}_{ij}}^r}  = \sum\limits_{j = 1}^K {{\eta _j}{P_{\rm{j}}}^t{d_{ij}}^{ - \alpha }} \begin{array}{*{20}{c}}
,&{i = 1,2, \ldots N}
\end{array},
\label{eq16}
\end{equation}
where $K$ is the number of radiation sources estimated in Section~\ref{MMPLD-based 3D Radiant Number Detection}.

We define ${\mathbf{\theta }}$ as the unknown vector of $5K$ parameters ${\bf{\theta }} = \left[ {{{\bf{\theta }}_1},{{\bf{\theta }}_2}, \ldots ,{{\bf{\theta }}_K}} \right],$ ${{\bf{\theta }}_j} = {\left[ {{\eta _j},x_j^t,y_j^t,z_j^t,{P_j}^t} \right]^T},j = 1,2, \cdots ,K$ and ${\bf{\Omega }}$ be the set of all acceptable values of ${\bf{\theta }}$. Given $K$, the cost function is established to estimate parameter, and the estimation of ${{\bf{\theta }}^{\bf{*}}}$ is a high dimensional optimization problem, as given by
\begin{equation}
\begin{array}{l}
{{\bf{\theta }}^{\bf{*}}} = \arg \mathop {\min }\limits_{\theta  \in {\bf{\Omega }}} sqrt\left( {{{\sum\limits_{i = 1}^N {\left( {{P_i}^r - {{\hat P}_i}^r} \right)} }^2}} \right)\\
= \arg \mathop {\min }\limits_{\theta  \in {\bf{\Omega }}} sqrt\left( {{{\sum\limits_{i = 1}^N {\left( {{P_i}^r - \sum\limits_{j = 1}^K {{\eta _j}{P_{\rm{j}}}^t{d_{ij}}^{ - \alpha }} } \right)} }^2}} \right),\\
{\rm{s}}{\rm{.t}}.\begin{array}{*{20}{c}}
{}&{{P_j}^t \ge 0,\begin{array}{*{20}{c}}
	{}&{j = 1,2, \cdots ,K}
	\end{array}}
\end{array},
\end{array}
\label{eq17}
\end{equation}
where ${P_i}^r,i = 1,2, \cdots ,N$ is the RSS of the $i$-th sampled position.

The algorithms can be used to solve this high dimensional parameters eatimation problem\cite{13}. SFLA is a swarm intelligence optimization algorithm based on post-heuristics. It is based on the evolution of individual memes and the global information exchange by memes. SFLA has many advantages, such as simple concept, few parameters, fast computing speed, strong global optimization ability, and easy implementation.

We develop SFLA to estimate the parameters ${{\bf{\theta }}^{\bf{*}}}$ in a $5K$ dimensional target search space. Firstly,  $P$ frogs (solutions) are randomly generated within ${\bf{\Omega }}$ to form the initial population, denoted as set ${\bf{\Lambda }} = \left\{ {{{\bf{F}}_1},{{\bf{F}}_2}, \ldots ,{{\bf{F}}_P}} \right\}$. The $s$-th frog represents a possible solution ${{\bf{F}}_s} = \left[ {{\bf{\theta }}_{s1}^*,{\bf{\theta }}_{s2}^*, \ldots ,{\bf{\theta }}_{sK}^*} \right]$ of the problem, where ${\bf{\theta }}_{sj}^* = {\left[ {{\eta _{sj}},x_{sj}^t,y_{sj}^t,z_{sj}^t,{P_{sj}}^t} \right]^T},j = 1,2, \cdots ,K$, is the parameter estimated of the $s$-th frog. Furthermore, the fitness function of the $s$-th frog can be defined according~\eqref{eq24}, given by 
\begin{equation}
\begin{array}{l}
\varphi (s) = sqrt\left( {{{\sum\limits_{i = 1}^N {\left( {{P_i}^r - \sum\limits_{j = 1}^K {{\eta _{sj}}{P_{{\rm{sj}}}}^t{d_{sij}}^{ - \alpha }} } \right)} }^2}} \right),\\
s = 1,2, \ldots ,P.
\end{array}
\label{eq18}
\end{equation}

Then we divide the whole population into $M$ subgroups in a descending order of $\varphi$. The frog ranked first is divided into the first subgroup, the frog ranked second is allocated into the second subgroup, the frog ranked $M$ is allocated into the $M$ subgroup, the  frog ranked $M + 1$ is allocated into the first subgroup, the frog ranked $M + 2$ is allocated into the second subgroup, and so on until all frogs were allocated.

We carry out local depth search for each subpopulation. Specifically, in each iteration of the subpopulation, we determine the frog with the worst $\varphi$ and the best $\varphi$ as ${{\bf{F}}_{nw}}$ and ${{\bf{F}}_{nb}}$, respectively. The frog with the best $\varphi$ in global population is denoted as ${{\bf{F}}_g}$. Only the worst individual ${{\bf{F}}_{nw}}$ in the current subpopulation is updated and the update strategy is 
\begin{equation}
\begin{array}{l}
{\varepsilon _{\rm{t}}} = \lambda  \cdot \left( {{{\bf{F}}_{nb}} - {{\bf{F}}_{nw}}} \right),\\
\left\| {{\varepsilon _{\min }}} \right\| < \left\| {{\varepsilon _t}} \right\| < \left\| {{\varepsilon _{\max }}} \right\|,
\end{array}
\label{eq19}
\end{equation}
where $\lambda $ is a random number between 0 and 1, and ${\varepsilon _t}$ denotes the distance that the frog moves, and $t = 1,2, \ldots ,Q$ in which $Q$ is the number of local iteration. ${\varepsilon _{\min }}$ and ${\varepsilon _{\max }}$ are the minimum and the maximum distances allowing frogs to move. The individual frog ${F_{nw}}$ position is updated by 
\begin{equation}
{{\bf{F}}_{nw}} = {{\bf{F}}_{nw}} + {\varepsilon _t}.
\label{eq20}
\end{equation}

If the updated frog is superior to the original frog in $\varphi$, the frog ${F_{nw}}$ of the original meme group can be replaced. Otherwise, ${{\bf{F}}_g}$ is used instead of ${{\bf{F}}_{nb}}$ to perform local location update, as given by
\begin{equation}
\begin{array}{l}
{\varepsilon _{\rm{t}}} = \lambda  \cdot \left( {{{\bf{F}}_g} - {{\bf{F}}_{nw}}} \right),\\
\left\| {{\varepsilon _{\min }}} \right\| < \left\| {{\varepsilon _t}} \right\| < \left\| {{\varepsilon _{\max }}} \right\|,
\end{array}
\label{eq21}
\end{equation}
\begin{equation}
{{\bf{F}}_{nw}} = {{\bf{F}}_{nw}} + {\varepsilon _t}.
\label{eq22}
\end{equation}

If the above steps still do not achieve a satisfactory fit, a new frog is randomly generated to replace the original ${{\bf{F}}_{nw}}$. When the local deep search of all subpopulations is completed, all frog individuals are remixed and sorted and the subpopulation is divided again. Then the local depth search is carried out again and the process is repeated until convergence conditions or the maximum number of iterations are met. The final output global optimal value ${{\bf{F}}_g}$ is the parameter ${{\bf{\theta }}^{\bf{*}}}$ we estimate.
\subsection{Propagation Model Self-Learning and Spectrum Reconstruction}
\label{Propagation}
In this section, we propose a model self-learning method to learn the electromagnetic environment. Considered that fading under the urban scenario can result in a loss of signal power without reducing the power of noise, it will cause performance degradation of a communication system. It can be solved with a semi-empirical propagation PL model to achieve an accurate estimate of the actual scenarios \cite{29}.

The RSS at the $i$-th point from the $j$-th radiation source can be modeled as
\begin{equation}
{P_{ij}}^r = {P_j}^t - {L_{ij}},
\label{eq23}
\end{equation}
where ${P_{ij}}^r$ and ${P_j}^t$ are the RSS at the $i$-th grid from the $j$-th radiation source and the transmitting power of the $j$-th radiation source, respectively. ${L_{ij}}$ is the PL between the $i$-th point and the $j$-th radiation source. RSS depends on the PL model and the transmitting power of all radiation sources in the measurement area. Then, the RSS ${P_i}^r$ of a certain point can be expressed as
\begin{equation}
\begin{array}{l}
{P_{ij}}^r\left[ {{\rm{mW}}} \right] = {10^{{P_{ij}}^r\left[ {{\rm{dBm}}} \right]/10}},\\
{P_i}^r\left[ {{\rm{mW}}} \right] = \sum\limits_{i = 1}^K {{P_{ij}}^r\left[ {{\rm{mW}}} \right]} ,
\end{array}
\label{eq24}
\end{equation}
where ${P_i}^r$ is the RSS value at the $i$-th point from all radiation sources.

If the semi-empirical propagation model is chosen properly, the awareness of the environment can be accurate by modifying the semi-empirical propagation model. It should be mentioned that for different scenarios, such as urban, suburban, indoor, mountain, etc, we can choose different semi-empirical models. In this paper, we choose an urban scenario in which the electromagnetic propagation characteristics of 3D space and the influence of shadow fading are considered. Besides, we reduce the impact of small-scale fading by averaging multiple RSS samples over time. Then the channel model is modeled according to the 3D airspace scenario, as given by \cite{14}
\begin{equation}
\begin{array}{l}
L = 32.4 + 20\log 10({f_{\rm{c}}})\\
+ 10(A \cdot {(h)^B})\log 10(d) + {\chi _\sigma },
\end{array}
\label{eq25}
\end{equation}
where $A$ and $B$ are environment-dependent parameters and they vary extraordinary in different scenarios. $H$ is the height relative to the ground. ${\chi _\sigma }$ is a zero-mean Gaussian variable with variance of $\sigma $ which represents the log-normal shadow fading.

According to~\eqref{eq3} and the data set we sampled, the parameters $A$, $B$ and $\sigma $ are estimated by solving a nonlinear fitting problem. For the $i$-th data, the actual measured path loss is
\begin{equation}
{L_i}^{{\rm{real}}} = \left( {10 \cdot \log 10\left( {\sum\limits_{j = 1}^K {{P_j}^t} } \right)} \right) - {P_i}^r,
\label{eq26}
\end{equation}
where ${P_j}$ and ${P_i}$ represents the transmitting power of the $j$-th radiation source and the RSS value of the $i$-th point.

Theoretical path loss can be obtained by ~\eqref{eq25}
\begin{equation}
\begin{array}{l}
{L_i}^{{\rm{theo}}} = \sum\limits_{j = 1}^K {{L_{ji}}} \\
= \sum\limits_{j = 1}^K {\left( {32.4 + 20\lg {f_c} + 10(A \cdot {{(h)}^B})\lg {d_{ij}} + {\chi _\sigma }} \right)} ,
\end{array}
\label{eq27}
\end{equation}
where ${d_{ij}}$ is the distance between the $i$-th point and the $j$-th radiation source. We define ${\bf{\Theta }} = \left[ {A,B,\sigma } \right]$ to be the unknown parameters of PL model. We obtain the estimated parameters ${\bf{\hat \Theta }}$ by constructing an optimization function, given by
\begin{equation}
{\bf{\hat \Theta }} = \arg \min {\left\| {{L_i}^{{\rm{theo}}} - {L_i}^{{\rm{real}}}} \right\|_2},
\label{eq28}
\end{equation}
where ${\bf{\hat \Theta }} = \left[ {\hat A,\hat B,\hat \sigma } \right]$ is the parameter to be estimated. There are only three parameters to be estimated here, so we can solve this problem by selecting the appropriate optimization algorithm.

After solving the ${\bf{\hat \Theta }}$, we can utilize ${\bf{\hat \Theta }}$ and ${{\bf{\theta }}^{\bf{*}}}$ to construct the 3D spectrum map. According to~\eqref{eq23} and~\eqref{eq24}, the RSS at any location can be evaluated by the parameters of radiation sources and the modified propagation model.
\section{SIMULATION RESULTS AND VALIDATIONS}
\label{SIMULATION}
\subsection{Experiment Setup}
\label{Experiment}
In this chapter, the performance of the proposed 3D spectrum map construction method is analyzed by simulation data. The ROI is a typical campus scenario. There are 9 buildings with the heights from 19 m to 35 m, and most of them are about 20 m. The scope of the entire region is $500{\kern 1pt} {\rm{m}} \times 500{\kern 1pt} {\rm{m}} \times 100{\kern 1pt} {\rm{m}}$, which means there are about 
	$7.7854 \times 10_{}^{ - 5}$ buildings per square meter. We discretize the space into a spectrum tensor of ${N_1} \times {N_2} \times {N_3} = 100 \times 100 \times 10$ with interval size of ${\rm{5{\kern 1pt} m}} \times {\rm{5{\kern 1pt} m}} \times {\rm{10{\kern 1pt} m}}$. It should be mentioned that it is impossible to acquire all spectrum data of the ROI in a real world. However, it is necessary for the performance evaluation of the spectrum map construction methods to acquire all spectrum data of the measurement area. In this case, we leverage the RT technique to obtain the 3D spectrum map simulation results of the ROI used for the performance evaluation. The RT technique has been widely used for radio propagation modeling, which has good performance for the small specific area and high frequency. In RT, the electromagnetic wave is viewed as a bunch of rays. When these rays interact with the scatterer surface and wedge, they can be reflected and diffracted. After all rays are tracked with the forward technique or the reverse technique, propagation parameters, i.e., electric field intensity or received signal strength of each ray, can be acquired. In this article, we only concentrate on the RSS. The simulation parameters of radiation sources are shown in Table\ref{tab1}. The values of other parameters are presented in Table\ref{tab2}. 
	
	We compare the results of the proposed method based on self-learning propagation model (SLPM) with that of the proposed method based on free space propagation model (FSPM). The results of IDW and High Accuracy Low Rank Tensor Completion (HaLRTC) are also provided. Root mean square error (RMSE), the Correct Detection Zone Ration (CDZR) and False Alarm Zone Ratios (FAZR) are selected as performance metrics. We also analysed the impact of radiation source number $K$. When comparing the performance of the spectrum map construction methods, we sample the data randomly, so that only a part of entries is retained in the spectrum tensor. The sampling rate indicates the percentage of the observed elements in the tensor.
\begin{table}[htbp]
	\centering
	\caption{The parameters of radiation sources for different number}
	\begin{tabular}{|c|ccc|c|}
		\hline
		\multirow{2}{*}{Number} & \multicolumn{3}{c|}{Location (m)}                            & \multirow{2}{*}{\begin{tabular}[c]{@{}c@{}}Transmission\\ power (w)\end{tabular}} \\ \cline{2-4}
		& \multicolumn{1}{c|}{x}   & \multicolumn{1}{c|}{y}    & z     &                                                                                   \\ \hline
		\multirow{2}{*}{2}      & \multicolumn{1}{c|}{310} & \multicolumn{1}{c|}{-239} & 2     & 1                                                                                 \\ \cline{2-5} 
		& \multicolumn{1}{c|}{235} & \multicolumn{1}{c|}{-105} & 2     & 1                                                                                 \\ \hline
		\multirow{3}{*}{3}      & \multicolumn{1}{c|}{345} & \multicolumn{1}{c|}{-365} & 33.77 & 1                                                                                 \\ \cline{2-5} 
		& \multicolumn{1}{c|}{205} & \multicolumn{1}{c|}{-265} & 2     & 1                                                                                 \\ \cline{2-5} 
		& \multicolumn{1}{c|}{245} & \multicolumn{1}{c|}{-95}  & 2     & 1                                                                                 \\ \hline
		\multirow{4}{*}{4}      & \multicolumn{1}{c|}{330} & \multicolumn{1}{c|}{-370} & 33.77 & 1                                                                                 \\ \cline{2-5} 
		& \multicolumn{1}{c|}{400} & \multicolumn{1}{c|}{-140} & 23.3  & 1                                                                                 \\ \cline{2-5} 
		& \multicolumn{1}{c|}{185} & \multicolumn{1}{c|}{-255} & 2     & 1                                                                                 \\ \cline{2-5} 
		& \multicolumn{1}{c|}{245} & \multicolumn{1}{c|}{-85}  & 2     & 1                                                                                 \\ \hline
	\end{tabular}
	\label{tab1}
\end{table}
\begin{table}[htbp]
	\centering
	\caption{Other simulation parameters}
	\begin{tabular}{|c|c|}
		\hline
		Parameter                                                               & Value                                                                                 \\ \hline
		Scenario                                                                & Urban                                                                                 \\ \hline
		\begin{tabular}[c]{@{}c@{}}The range of\\ measurement area\end{tabular} & \begin{tabular}[c]{@{}c@{}}X(m): 0-500 \\ Y(m): -450-50   \\ Z(m): 0-100\end{tabular} \\ \hline
		Frequency                                                               & 100 MHz                                                                                \\ \hline
		Antenna type                                                            & Omnidirectional                                                                       \\ \hline
	\end{tabular}
	\label{tab2}
\end{table}
\subsection{Comparison and Validation}
\label{Comparison}
$i)$ \textit{Comparison of Received Signal Strength Recovery:} In this section, the RMSE are chosen as the performance evaluation criterion of RSS recovery average error. The average error is defined as the difference of each grid point in RSS between the estimated REM and the real REM, given by
\begin{equation}
{\rm{RMSE}} = \frac{1}{N}\sum\limits_{i = 1}^N {\sqrt {{{\left| {\frac{{P_i^{^r{\rm{est}}} - P_i^{^r{\rm{real}}}}}{{P_i^{^r{\rm{real}}}}}} \right|}^2}} },
\label{eq29}
\end{equation}
where $P_i^{^r{\rm{est}}}$ and $P_i^{^r{\rm{real}}}$ are the estimated value and the actual value respectively. $N = {N_1} \times {N_2} \times {N_3}$ is the total number of grids.

In Figure~\ref{fig4}, we compare the results of the proposed method based on SLPM with that of the proposed method based on FSPM. The results of IDW and HaLRTC are also provided. Given the sampling rate $r$, the sampling locations are arbitrarily chosen from the measurement area. It shows that the RMSE of RSS recovery of all algorithms decrease with the increase of sampling rate. The proposed method based on SLPM outperforms other three algorithms. The data-driven method IDW and HaLRTC has poor performance in the complex urban environment. In addition, the proposed method based on the SLPM can better guarantee the recovery of the actual electromagnetic environment than the method directly based on FSPM. 
	
	We also compare the influence of the different number $K$ of radiation sources on the RSS recovery errors of the four algorithms with the same sampling rate. The results can be seen in Figure~\ref{fig5}. We can see that the RMSE of the RSS recovery error of all algorithms decrease with the decrease of radiation source number. The recovery performance of the data-driven methods, i.e., IDW and HaLRTC, is far inferior to the two model-driven methods when multi-radiation sources exist. This is because the data-driven methods have high requirements on the number and locations of sampled data. Moreover, the spectrum recovery performance of the plane with unsampled height is poor but the sampling rate is high. By comparing the proposed method based on the SLPM with that based on the FSPM, self-learning of the model can further improve the recovery performance. 

Figures~\ref{fig4} and \ref{fig5} also show that the construction accuracy of model-driven spectrum map methods is higher than that of data-driven spectrum map methods in terms of the sampling rate and the radiation source number. In addition, although the proposed algorithm based on the FSPM and the SLPM respectively have similar convergence rates, the recovery accuracy of the latter is high.
\begin{figure}[htbp]
	\centering
	\includegraphics[width=0.5\textwidth]{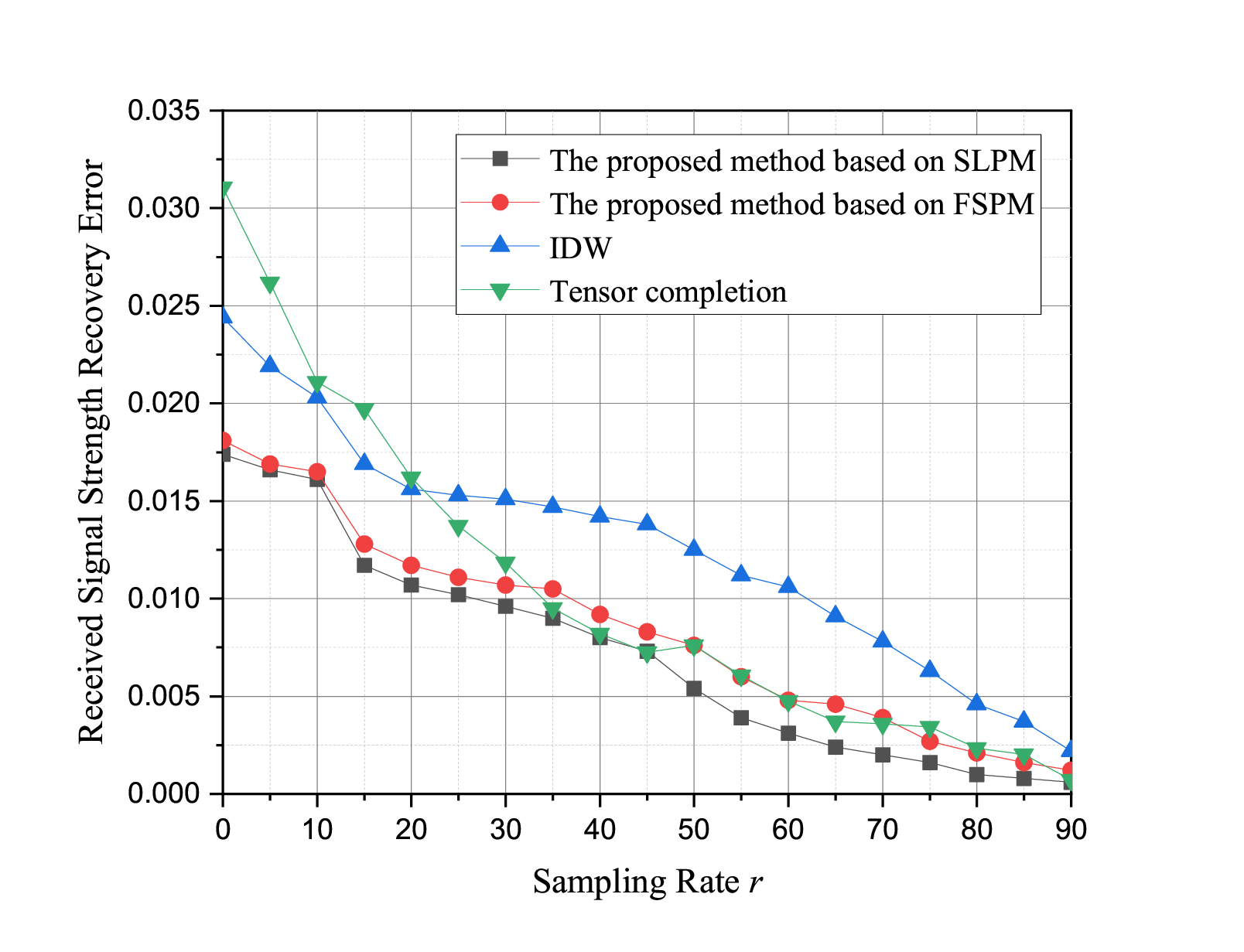}
	\caption{The RMSE of RSS recovery error vs. the sampling rate $r$ ($K$=3).}
	\label{fig4}
\end{figure}
\begin{figure}[htbp]
	\centering
	\includegraphics[width=0.5\textwidth]{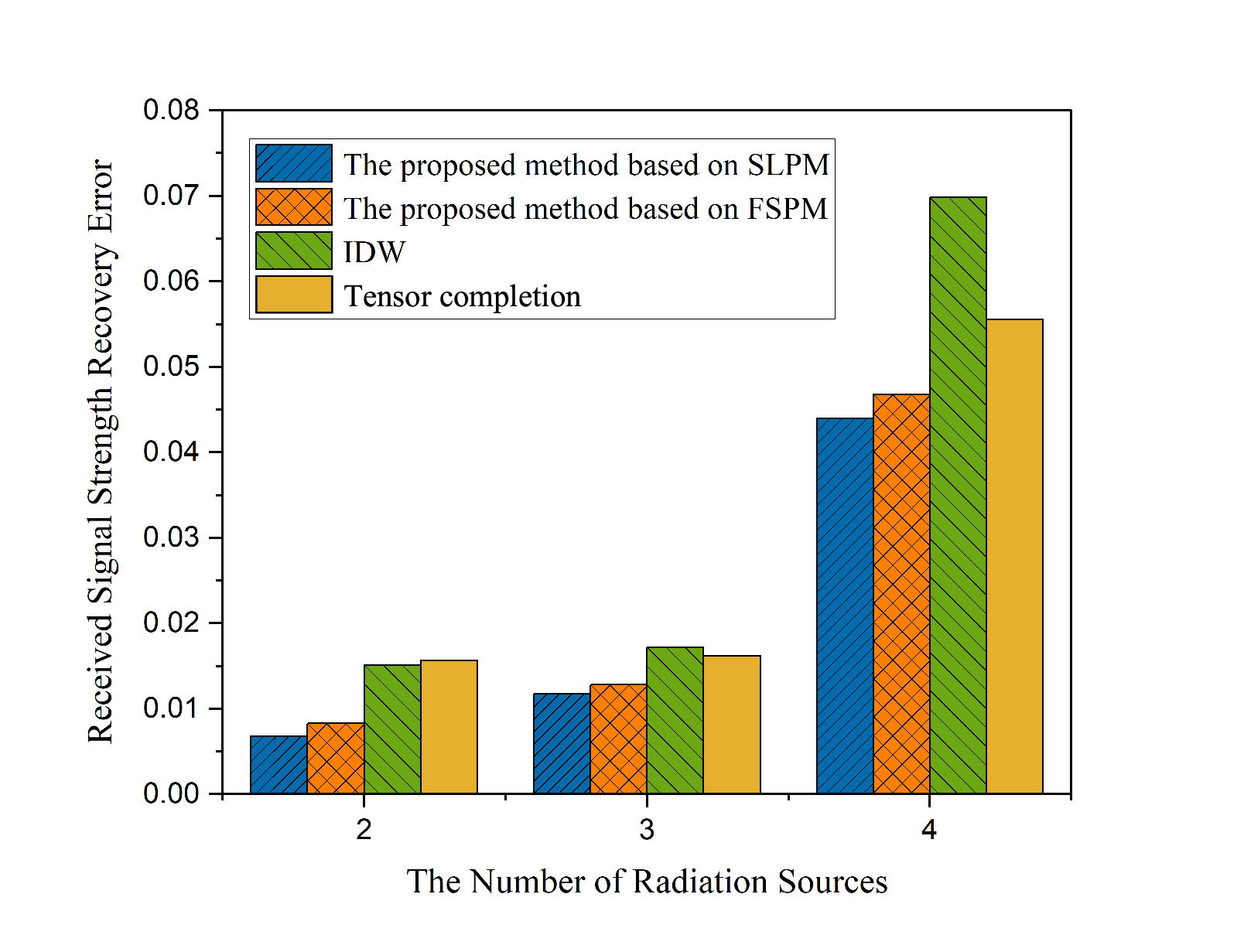}
	\caption{ The RMSE of RSS recovery error vs. the radiation source number $K$ ($r$=0.2).}
	\label{fig5}
\end{figure}

$ii)$ \textit{Comparison of CDZR and FAZR:} In $i)$, we calculated the RMSE of RSS recovery results between the real spectrum map and the estimated spectrum map to quantify the similarity. To determine where spectrum opportunities are lost and where CR conflicts with ongoing transmissions of radiation sources in the region, the Correct Detection Zone Ration (CDZR) and False Alarm Zone Ratios (FAZR) are introduced in \cite{10}. 

\begin{figure}[htbp]
	\centering
	\includegraphics[width=0.5\textwidth]{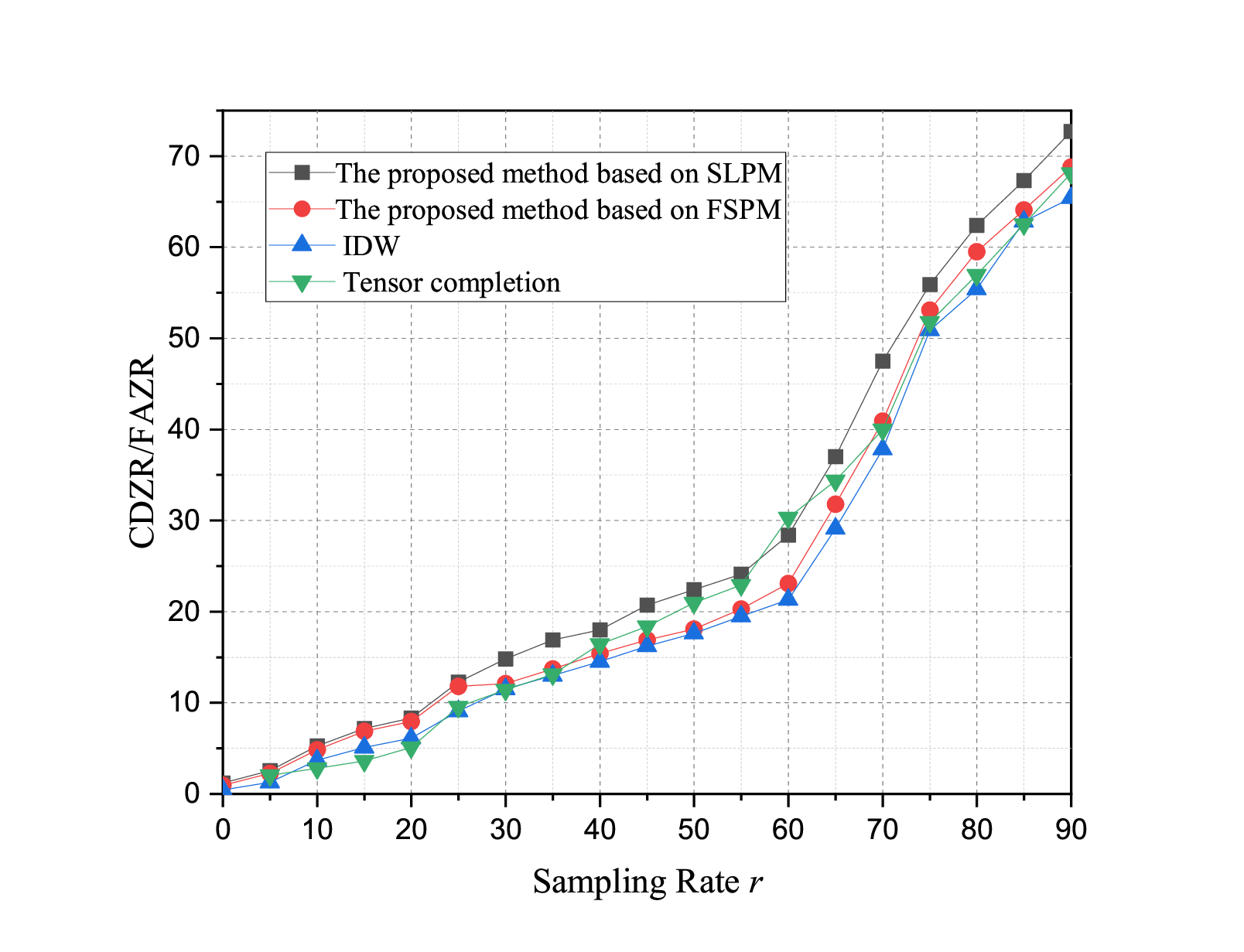}
	\caption{The CDZR/FAZR of RSS recovery error vs. the sampling rate $r$ ($K$=3).}
	\label{fig6}
\end{figure}
The intersection of the zones of true REM and estimated REM defines the zones that are correctly and incorrectly determined. Regions where both the estimated spectrum map and the actual spectrum map correctly identify forbidden is marked as the Correct Detection Zone of type 1 ($Z_1^{{\rm{C}}{{\rm{D}}_j}}$), while identifying permitted is marked as Correct Detection Zone of type 0 ($Z_0^{{\rm{C}}{{\rm{D}}_j}}$). False Alarm Zone $Z_0^{{\rm{F}}{{\rm{A}}_j}}$ represents the area where communication is allowed but is incorrectly detected as the prohibitive zone. Missed Detection Zone $Z_1^{{\rm{M}}{{\rm{D}}_j}}$ represents the area where transmission is forbidden but missed detection. The CDZR and FAZR can be obtained from the following equations
\begin{equation}
\begin{array}{l}
{\rm{FAZ}}{{\rm{R}}_j} = \frac{{S\left( {Z_0^{{\rm{F}}{{\rm{A}}_j}}} \right)}}{{S\left( {Z_0^{{\rm{F}}{{\rm{A}}_j}}} \right) + S\left( {Z_0^{{\rm{C}}{{\rm{D}}_j}}} \right)}}\begin{array}{*{20}{c}}
,&{j = 1,2, \ldots ,K}
\end{array},\\
{\rm{FAZR}} = \sum\limits_{j = 1}^K {{\rm{FAZ}}{{\rm{R}}_j}} ,
\end{array}
\label{eq30}
\end{equation}
\begin{equation}
\begin{array}{l}
{\rm{CDZ}}{{\rm{R}}_j} = \frac{{S\left( {Z_1^{{\rm{C}}{{\rm{D}}_j}}} \right)}}{{S\left( {Z_1^{{\rm{C}}{{\rm{D}}_j}}} \right) + S\left( {Z_1^{{\rm{M}}{{\rm{D}}_j}}} \right)}}\begin{array}{*{20}{c}}
,&{j = 1,2, \ldots ,K}
\end{array},\\
{\rm{CDZR}} = \sum\limits_{j = 1}^K {{\rm{CDZ}}{{\rm{R}}_j}} ,
\end{array}
\label{eq31}
\end{equation}
where $S\left( Z \right)$ is the size of the zone $Z$. ${\rm{FAZ}}{{\rm{R}}_j}$ and ${\rm{CDZ}}{{\rm{R}}_j}$ are Correct Detection Zone Ration and False Alarm Zone Ratios at the $j - th$ radiation source respectively. Having higher CDZR means disturbing the active user is less probable. Having less FAZR means white spaces can be utilized better. Better spectrum map construction performance fit better on the true spectrum map contour with predetermined threshold value.

We compared the ${\rm{CDZR}}/{\rm{FAZR}}$ curves as shown in Figure~\ref{fig6}. The simulation results show that based on the knowledge of radiation sources and self-learning of the model, the proposed algorithm can make better use of the white spaces and reduce the interference to the radiation sources than the other two algorithms. Figure~\ref{fig6} can also be seen that the data-driven methods cannot achieve satisfactory performance, since their performance largely depends on the sampling positions. The data of unsampled position, which is interpolated by the data-driven methods, is smaller than the maximum value of the sampled data. Then if the sampling position is far away from the radiation source, it is difficult to recover its nearby real spectrum data.
\begin{figure}[htbp]
	\centering
	\includegraphics[width=0.5\textwidth]{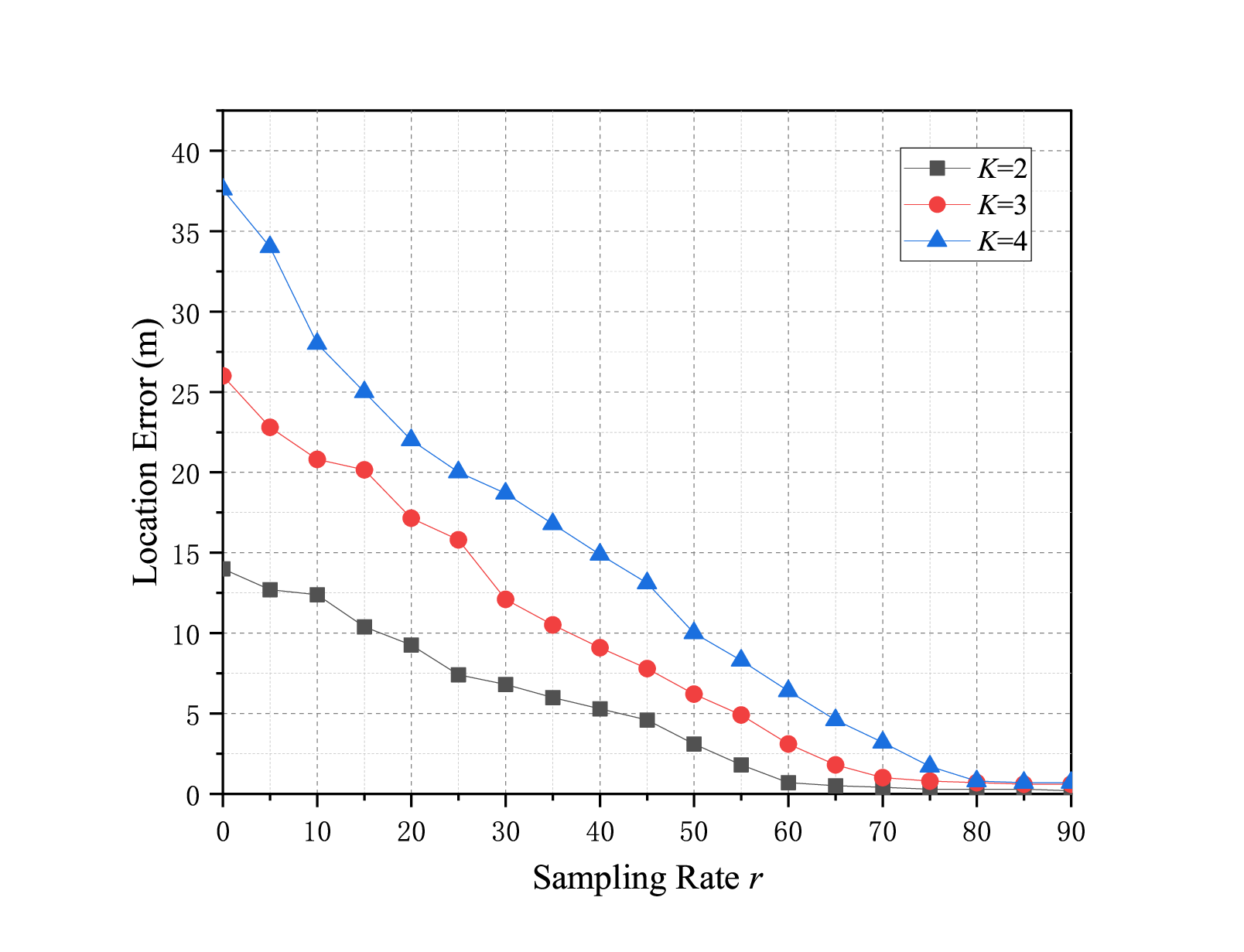}
	\caption{The localization errors vs. the sampling rate $r$.}
	\label{fig7}
\end{figure}
\begin{figure}[htbp]
	\centering
	\includegraphics[width=0.5\textwidth]{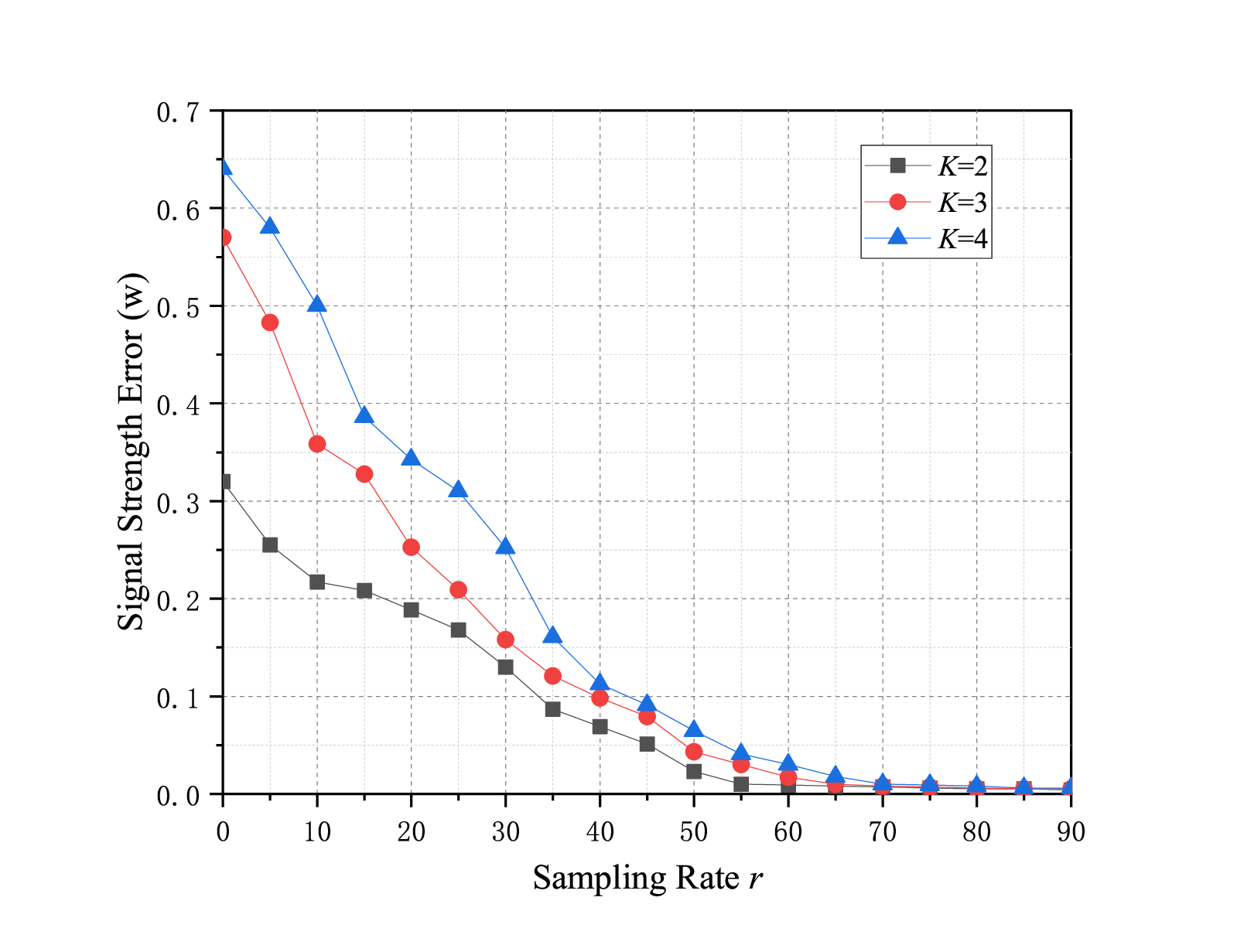}
	\caption{The source signal strength recovery errors vs. the sampling rate $r$.}
	\label{fig8}
\end{figure}
\begin{figure}[htbp]
	\centering
	\includegraphics[width=0.5\textwidth]{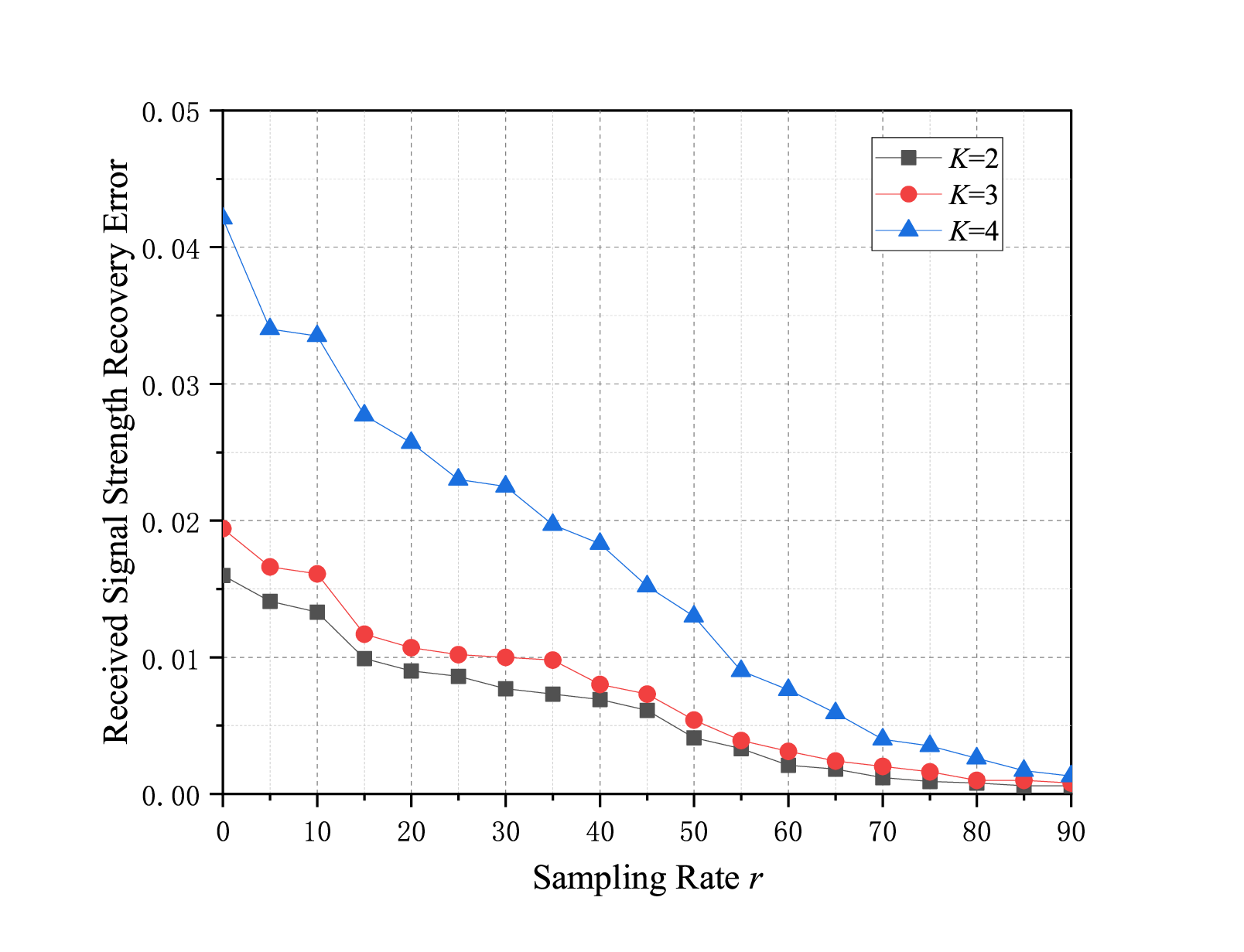}
	\caption{The RSS recovery errors vs. the sampling rate $r$.}
	\label{fig9}
\end{figure}
\begin{figure}[htbp]
	\centering
	\includegraphics[width=0.5\textwidth]{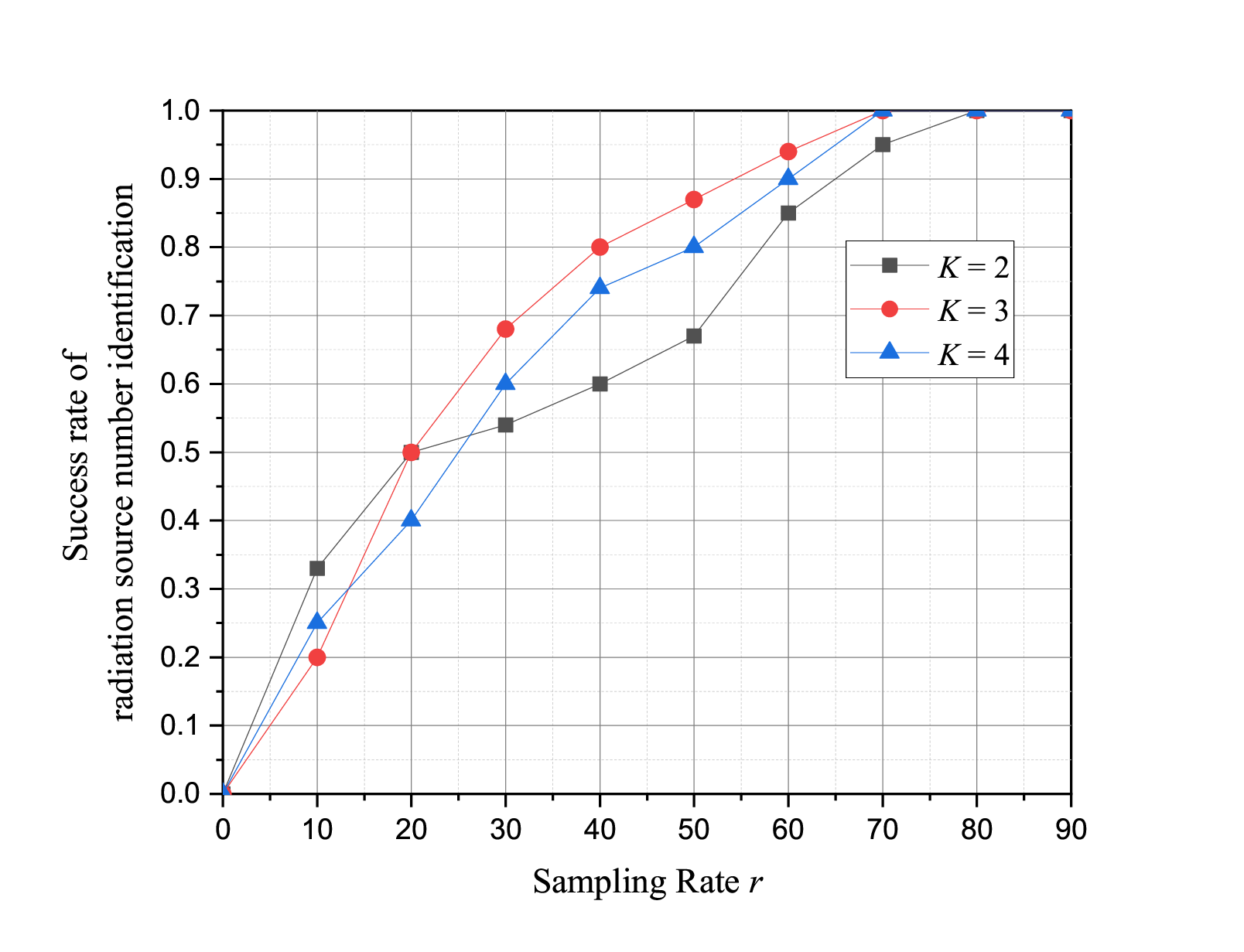}
	\caption{The success rate of radiation source number identification vs. the sampling rate $r$.}
	\label{fig10}
\end{figure}
\begin{figure} 
	\centering
	\vspace{-0.6cm}
	\begin{minipage}{0.48\linewidth}		
		\centerline{\includegraphics[width=4.5cm]{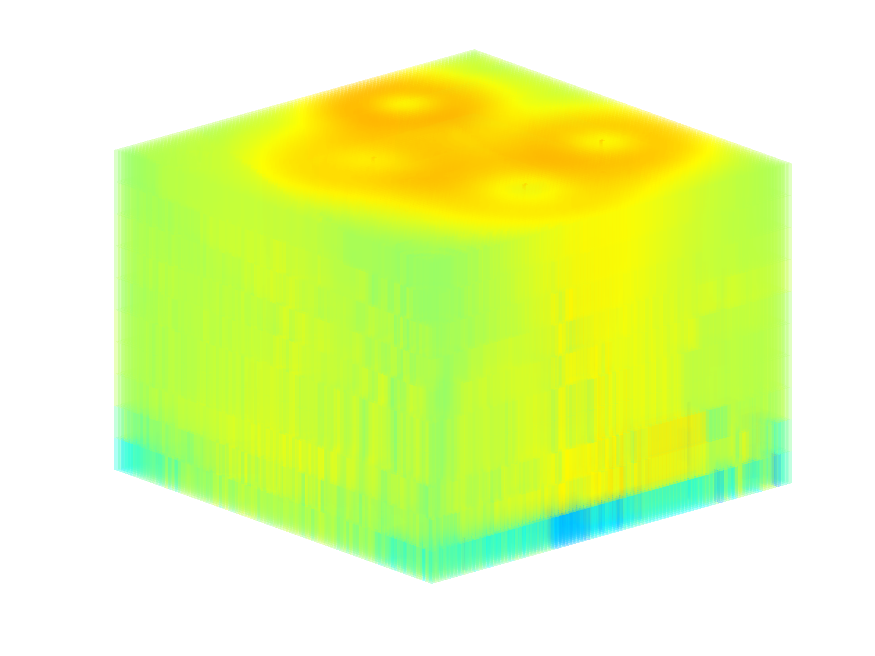}}		
		\centerline{(a)}		
	\end{minipage}	
	\vfill	
	\begin{minipage}{.48\linewidth}		
		\centerline{\includegraphics[width=4.5cm]{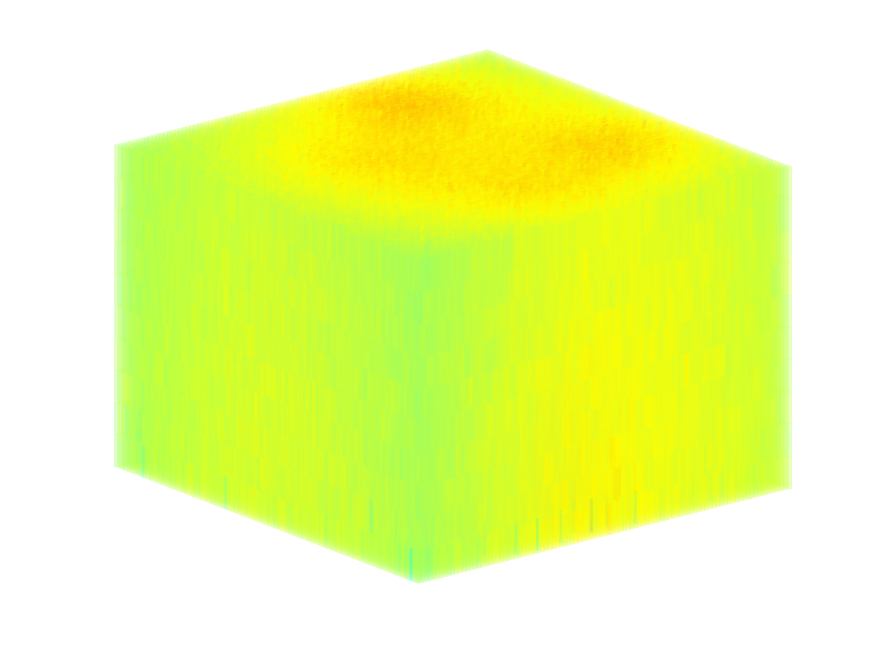}}		
		\centerline{(b)}		
	\end{minipage}	
	\hfill	
	\begin{minipage}{0.48\linewidth}		
		\centerline{\includegraphics[width=4.5cm]{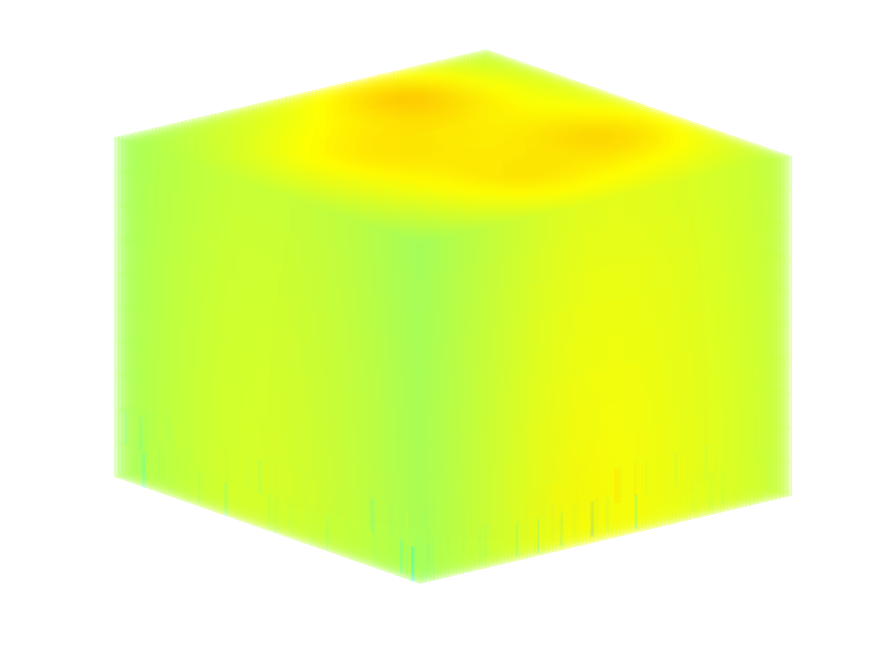}}		
		\centerline{(c)}		
	\end{minipage}	
	\vfill	
	\begin{minipage}{0.48\linewidth}		
		\centerline{\includegraphics[width=4.5cm]{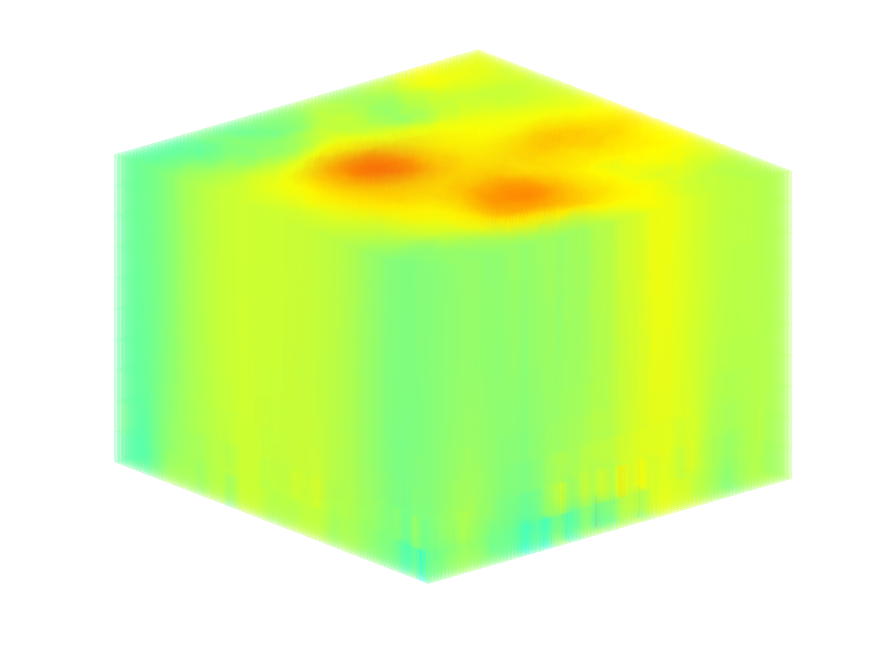}}		
		\centerline{(d)}		
	\end{minipage}	
	\hfill	
	\begin{minipage}{0.48\linewidth}		
		\centerline{\includegraphics[width=4.5cm]{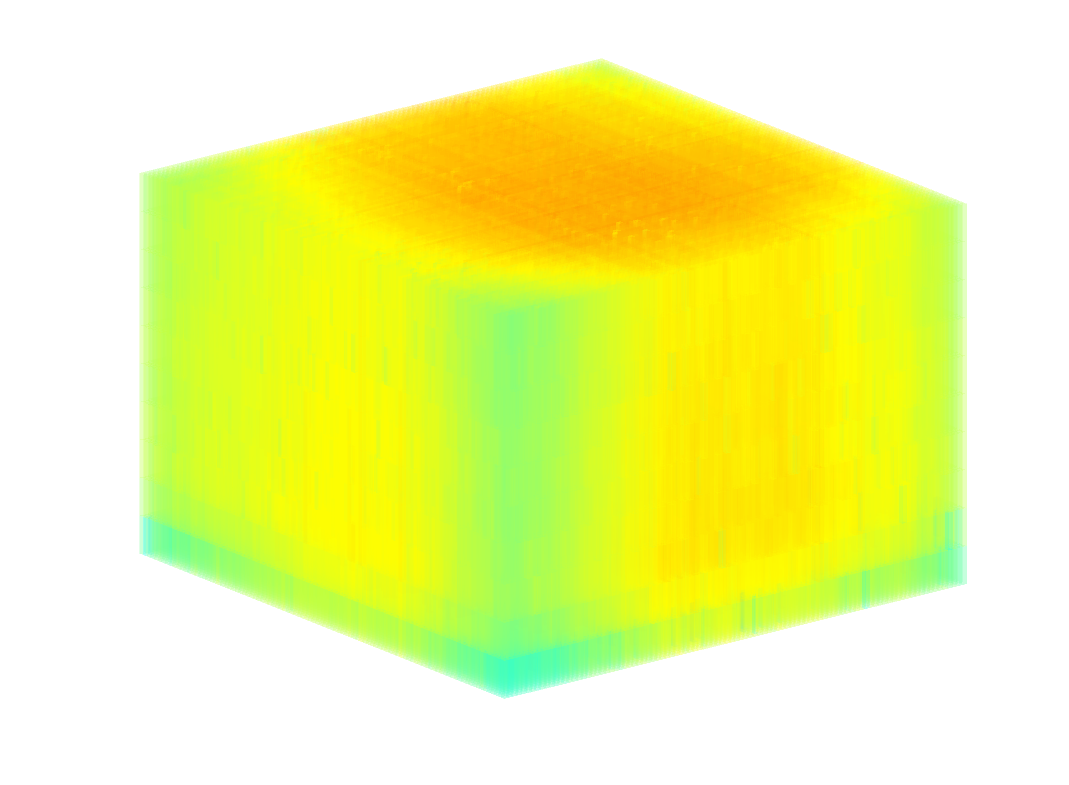}}		
		\centerline{(e)}		
	\end{minipage}	
	\caption{3D visualization results of spectrum situation recovery (K=4). (a): the simulation of spectrum map. (b)-(e): the spectrum situation recovery results of the proposed method based on the SLPM, FSPM, IDW and Tensor completion, respectively.}	
	\label{fig11}	
\end{figure}

$iii)$ \textit{Impact of Radiation Source Number:} We study the impact of radiation source number $K$ on the performance of the proposed algorithm including localizations, signal strength recovery, the success rate of radiation source number identification and RSS recovery.

The average localization error is defined as 
\begin{equation}
{\rm{LOC\_E}} = \frac{1}{K}\sum\limits_{i = 1}^K {{{\left\| {{\rm{lo}}{{\rm{c}}_i}^{{\rm{est}}} - {\rm{lo}}{{\rm{c}}_i}^{{\rm{real}}}} \right\|}_2}},
\label{eq32}
\end{equation}
where ${\rm{lo}}{{\rm{c}}_i}^{{\rm{est}}}$ and ${\rm{lo}}{{\rm{c}}_i}^{{\rm{real}}}$ denote the estimated and the real locations of radiation sources in Cartesian coordinate. 

The signal strength recovery error is defined as
\begin{equation}
{\rm{SS\_E}} = \frac{1}{K}\sum\limits_{i = 1}^K {{{\left\| {{\rm{si}}{{\rm{g}}_i}^{{\rm{est}}} - {\rm{si}}{{\rm{g}}_i}^{{\rm{real}}}} \right\|}_2}},
\label{eq33}
\end{equation}
where ${\rm{si}}{{\rm{g}}_i}^{{\rm{est}}}$ and ${\rm{si}}{{\rm{g}}_i}^{{\rm{real}}}$ represent estimated and true signal strengths of radiation sources. 

In Figures~\ref{fig7} and \ref{fig8}, the impact of $K$ on the performances of localizations and signal strength recovery is analyzed. It can be seen that the performances of three algorithms are all improved as $K$ decreases. The localization error and signal strength recovery error decrease when the number of radiation sources is reduced. These factors are the key to determine the performance of the proposed algorithm. The RMSE of RSS recovery also decreases with the increase of $K$, as shown in Figure~\ref{fig9}. This is because increasing $K$ will aggravate the computation and lead to more interactions between radiation sources. In addition, when the sampling rate $r$ grows, the performances of these four algorithms improve significantly. 

\sethlcolor{yellow}
Furthermore, we analyze the success rate of radiation source number identification as shown in Figures~\ref{fig10}. It can be seen that when the sampling rate is over 0.3, our MMPLD clustering algorithm can identify the radiation source number with the success rate of 0.7.

$iv)$ \textit{3D Spectrum Map Visualization:} Figures~\ref{fig11} and \ref{fig12} present the 3D spectrum map visualization results of the IDW, the tensor completion, the proposed method based on the SLPM and FSPM. In order to analyze the construction situation of spectrum map clearly, we observe two figures from different angles.
	
	Figure~\ref{fig11} shows the 3D visualization of spectrum situation recovery. By comparing Figure~\ref{fig11} (b)-(e), we can see that the IDW and tensor completion perform poorly in 3D spectrum situation recovery, as their performance only depends on the data. Figure~\ref{fig11} (b) and (c) can be seen that the proposed method based on the SLPM can recover the original spectrum situation veritably with the consideration of shadow fading in the urban environment.
	
	Figure~\ref{fig12} (b)-(e) further visually reveal that our proposed method performs well in the spectrum situation recovery at radiation sources and in the source signal strength recovery of radiation sources. As shown in Figure~\ref{fig12} (b) and (c), by exploiting the knowledge of radiation sources, the spectrum situation can be well recovered around the radiation sources. Besides, considering the self-learning of the propagation model, the visualization in Figure~\ref{fig12}(b) can capture the shadowing, which is superior than the counterpart in Figure~\ref{fig12} (c). The IDW and tensor completion in Figure~\ref{fig12} (d) and (e) failed to recover the spectrum data around the radiation sources due to the influence of sampling positions  . It can be seen that the construction performance of the tensor completion method is very poor when the sampling data are relatively unevenly distributed in the ROI.
\begin{figure} 
	\centering
	\begin{minipage}{0.48\linewidth}		
		\centerline{\includegraphics[width=4.5cm]{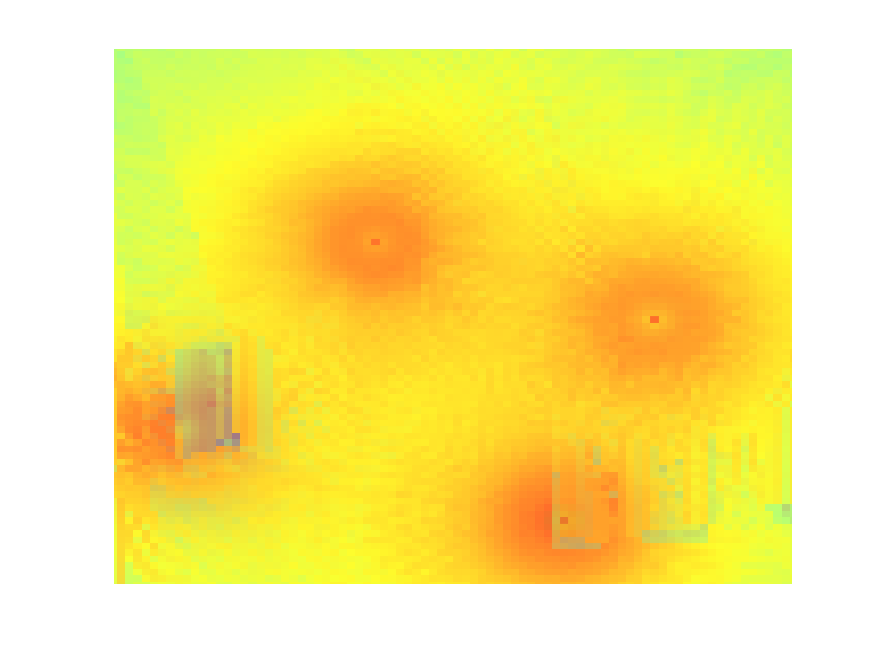}}		
		\centerline{(a)}		
	\end{minipage}	
	\vfill	
	\begin{minipage}{.48\linewidth}		
		\centerline{\includegraphics[width=4.5cm]{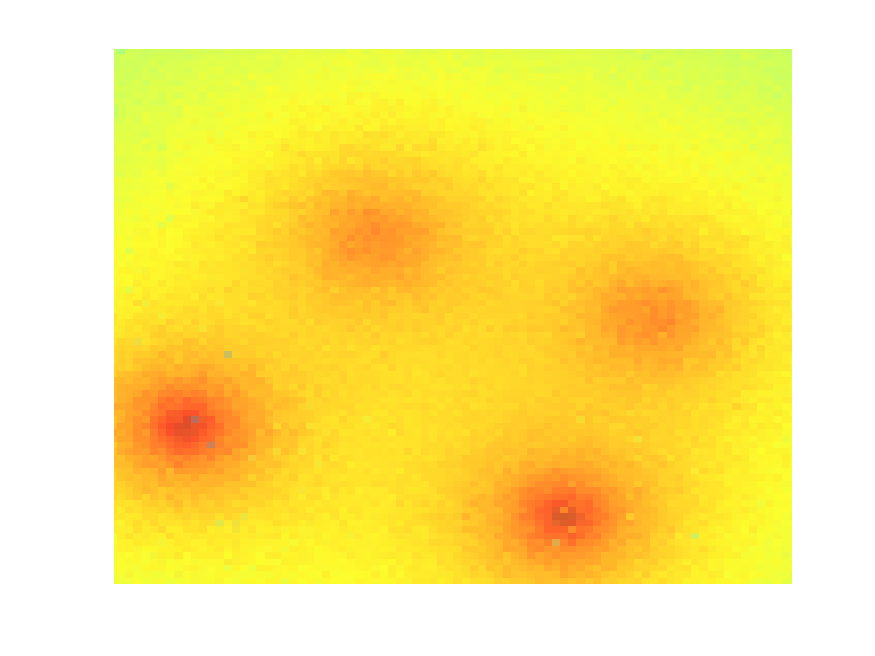}}		
		\centerline{(b)}		
	\end{minipage}	
	\hfill	
	\begin{minipage}{0.48\linewidth}		
		\centerline{\includegraphics[width=4.5cm]{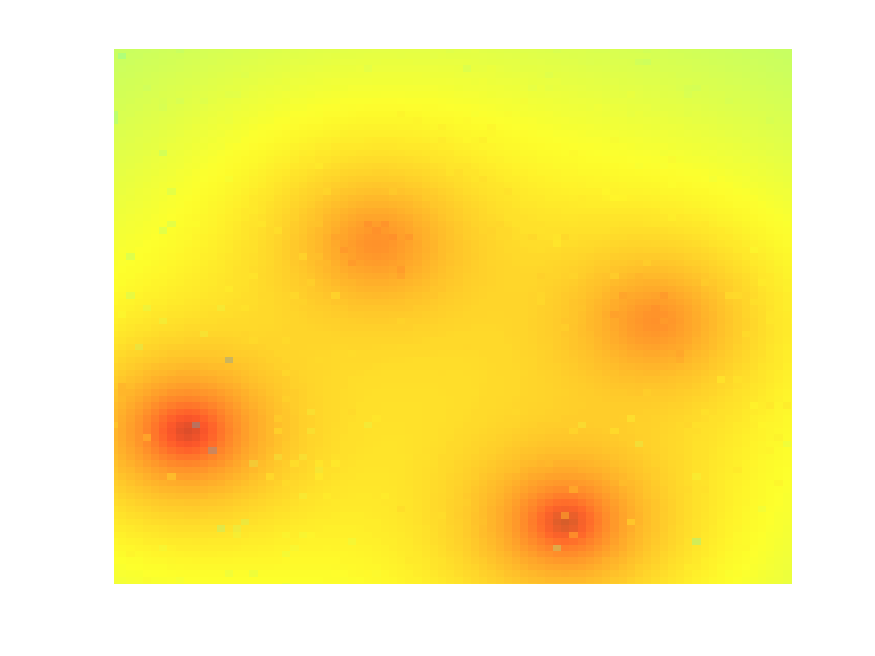}}		
		\centerline{(c)}		
	\end{minipage}	
	\vfill	
	\begin{minipage}{0.48\linewidth}		
		\centerline{\includegraphics[width=4.5cm]{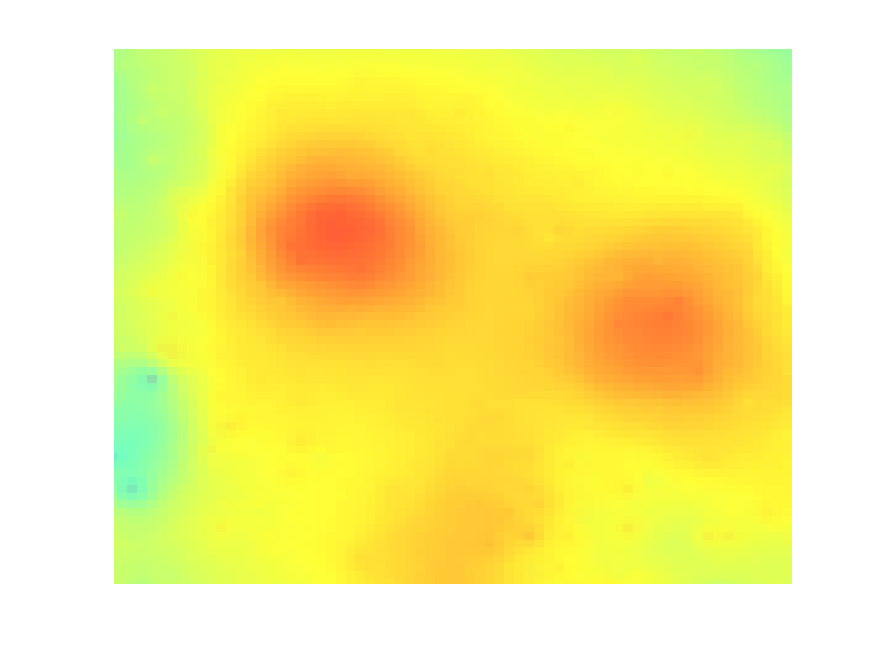}}		
		\centerline{(d)}		
	\end{minipage}	
   \hfill	
   \begin{minipage}{0.48\linewidth}		
     	\centerline{\includegraphics[width=4.5cm]{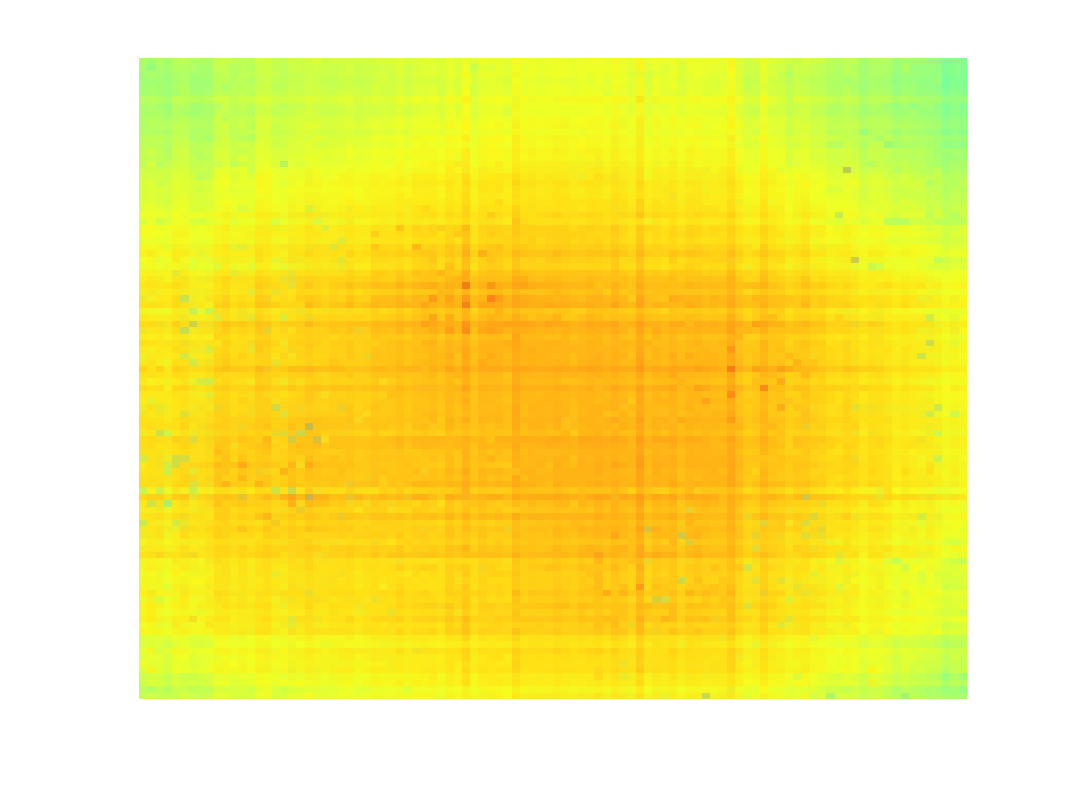}}		
    	\centerline{(e)}		
   \end{minipage}	
	\caption{Visualization results of spectrum situation recovery at radiation sources (K=4). (a): the simulation of spectrum map. (b)-(e): the spectrum situation recovery at radiation sources of the proposed method based on the SLPM, FSPM, IDW and Tensor completion, respectively.}	
	\label{fig12}	
\end{figure}
\section{CONCLUSION}
\label{CONCLUSION}
In this paper, we have investigated 3D spectrum mapping under multi-radiation source urban scenarios. We have constructed the spectrum map indirectly through the study of the parameters of radiation sources and the environment of the 3D space. We have proposed the 3D spectrum mapping and reconstruction scheme, which is composed of an extraction of radiation source knowledge and self-learning of the propagation model. We have designed a MMPLD-based clustering algorithm to detect the number of radiation sources and a SFLA-based optimizing estimation algorithm to locate the radiation sources. Propagation model self-learning based on sampling data has also been proposed, which can describe the real time electromagnetic propagation states. We have developed the empirical PL model by sampling data to make the proposed algorithm self learn. We have compared the RMSE of RSS recovery and CDZR/FAZR curves among four algorithms i.e., the proposed method based on the SLPM, FSPM, IDW and tensor completion. Furthermore, the impact of radiation source number has also been studied. Simulation results have shown that the proposed method has obvious advantages in the RSS recovery under multi-radiation source urban scenarios. Further research will be conducted to sampling optimization and dynamic reconstruction of radiation source movement.
\section*{ACKNOWLEDGEMENT}
\label{ACKNOWLEDGEMENT}
This work was supported in part by the National Key Scientific Instrument and Equipment Development Project under Grant No. 61827801, in part by the open research fund of State Key Laboratory of Integrated Services Networks, No. ISN22-11, in part by Natural Science Foundation of Jiangsu Province, No. BK20211182, and in part by the open research fund of National Mobile Communications Research Laboratory, Southeast University, No. 2022D04.

\bibliographystyle{IEEEtran}
\bibliography{myref}
\biographies
\begin{CCJNLbiography}{wangjie.eps}{Jie Wang}
received the B.S. degree in internet of things engineering from the
College of Information Science and Technology, Nanjing Forestry University of China, Nanjing, China, in 2021. She is currently pursuing the Ph.D. degree in communications and information systems with the College of Electronic and Information Engineering, Nanjing University of Aeronautics and Astronautics. Her current research interests is spectrum mapping.
\end{CCJNLbiography}
\begin{CCJNLbiography}{ZhiPengLin.eps}{Zhipeng Lin}
received the Ph.D. degrees from the School of Information and Communication Engineering, Beijing University of Posts and Telecommunications, Beijing, China, and the School of Electrical and Data Engineering, University of Technology of Sydney, NSW, Australia, in 2021. He is currently an Associate Researcher in the College of Electronic and Information Engineering, Nanjing University of Aeronautics and Astronautics, Nanjing, China. His current research interests include signal processing, massive MIMO, spectrum sensing, and UAV communications.
\end{CCJNLbiography}
\begin{CCJNLbiography}{zhuqiuming.eps}{Qiuming Zhu}
received the B.S. degree in electronic engineering and the M.S. and
Ph.D. degrees in communication and information system from the Nanjing University of Aeronautics and Astronautics (NUAA), Nanjing, China,
in 2002, 2005, and 2012, respectively. He is currently a Professor in the College of Electronic and Information Engineering, Nanjing University of Aeronautics and Astronautics, Nanjing, China. His current research interests include channel sounding, modeling, and emulation for the fifth/sixth generation (5G/6G) mobile communication, vehicle-to-vehicle (V2V) communication, and unmanned aerial vehicles (UAV) communication systems.
\end{CCJNLbiography}
\begin{CCJNLbiography}{wuqihui.eps}{Qihui Wu}
received the B.S. degree in communications engineering and the M.S.
and Ph.D. degrees in communications and information system from the PLA University of Science and Technology, Nanjing, China, in 1994, 1997, and 2000, respectively. He is currently a Professor with the College of Electronic and Information Engineering, Nanjing University of Aeronautics and Astronautics. His current research interests include algorithms and optimization for cognitive wireless networks, soft-defined radio, and wireless communication systems.
\end{CCJNLbiography}
\begin{CCJNLbiography}{lantianxu.eps}{Tianxu Lan}
 received the B.S. degree in information engineering from Nanjing University of Aeronautics and Astronautics (NUAA) in 2019. He is currently working towards the master degree in electronic information engineering, NUAA. His current research is reconstruction of spectrum situation.	
\end{CCJNLbiography}
\begin{CCJNLbiography}{zhaoyi.eps}{Yi Zhao}
received the B.S. degree in information engineering from Nanjing University of Aeronautics and Astronautics (NUAA) in 2021. He is currently working towards the master degree in electronic information engineering, NUAA. His current research is temporal and spatial prediction and reconstruction of spectrum situation.	
\end{CCJNLbiography}
\begin{CCJNLbiography}{YunPengBai.eps}{Yunpeng Bai}
received the B.E. degree in electrical engineering and automation from Shandong University (SDU) in 2019. He is currently working towards the master degree in electronic information engineering, NUAA. His current research is UAV path planning.
\end{CCJNLbiography}
\begin{CCJNLbiography}{zhongweizhi.eps}{Weizhi Zhong}
received her B.S. and M.S. degrees in communication and information system from the Jilin University, and received the Ph.D. degree in communication and information system from the Harbin Institute of Technology, respectively. She is an associate professor in the College of Astronautics, Nanjing University of Aeronautics and Astronautics (NUAA). Her research interests include millimeter wave communication for the fifth generation (5G), massive MIMO technique and beamforming and beam tracking technique. 
\end{CCJNLbiography}

\end{document}